\documentclass[notitlepage,letterpaper]{article}
\usepackage[ansinew]{inputenc} 
\usepackage{graphicx}
\usepackage{graphicx}
\usepackage{amsfonts}
\usepackage{amssymb}
\usepackage{color}
\usepackage{harvard}
\usepackage{cite}
\usepackage{geometry}      
\usepackage{epstopdf}
\usepackage{graphicx,color}
\usepackage{subfigure} 

\begin{document}

\title{Quasi-static thermal  evolution of compact objects}

\author{
\textbf{L. Becerra} \\
\textit{Escuela de F\'isica, Facultad de Ciencias,} \\ \textit{Universidad Industrial de Santander, Bucaramanga 680002, Colombia}; \\
\textbf{ H. Hernández} \\
\textit{Laboratorio de F\'{\i}sica Te\'{o}rica, Departamento de F\'{\i}sica, Facultad de Ciencias, }\\ 
\textit{Universidad de Los Andes,  M\'{e}rida 5101, Venezuela.} \\
\textbf{L. A. Núñez} \\
\textit{Escuela de F\'isica, Facultad de Ciencias,} \\ 
\textit{Universidad Industrial de Santander, Bucaramanga 680002, Colombia} and \\
\textit{Centro de F\'{\i}sica Fundamental, Departamento de F\'{\i}sica, Facultad de Ciencias}, \\
\textit{Universidad de Los Andes, M\'{e}rida 5101, Venezuela.}
}
\maketitle

\begin{abstract}
We study under what conditions the thermal peeling is present for dissipative local and quasi-local anisotropic spherical matter configurations. The thermal peeling occurs when different signs in the velocity of fluid elements appears, giving rise to the splitting of the matter configuration.  The evolution is considered in the quasi-static approximation and the matter contents are radiant, anisotropic (unequal stresses) spherical local and quasi-local fluids.  The heat flux and the associated temperature profiles are described by causal thermodynamics consistent with this approximation. 
It is found some particular, local and quasi-local equation of state for ultra-dense matter configurations exhibit thermal peeling when most of the radiated energy is concentrated at the middle of the distribution. This effect, which appears to be associated with extreme astrophysical scenarios (highly relativistic and very luminous gravitational system expelling its outer mass shells), is very sensible to energy flux profile and to the shape of the luminosity emitted by the compact object.
\end{abstract}

\section{Introduction}
Neutron stars, emerging as relics of extreme events, appear to be the observable densest cold matter objects in our Universe. These very fascinating entities cool through a blend complex mechanisms combining neutrino transport from their interiors and the emission of photons from their surfaces. Neutrinos dominate energy losses during the first $\sim 10^{5}$ years after the birth of these objects while photons are responsible for the observed thermal emissions detected from neutron stars older than hundreds of years. The possibility to observe of these objects depends  on the overall rate of neutrino leaked from the object interior and photon emitted from it surface.  With the significant amount of data that have been accumulated out from Chandra, Newton X-ray and Hubble space missions, we are entering in an era with the possibility to outline more precisely the mechanisms of the thermal behavior from isolate neutron stars (see \cite{PageReddy2012,TsurutaEtal2009} and references therein). 

Despite this very detailed emerging modelling on the microphysics of the neutron star cooling (neutrino transport/emission processes and superfluidity of the constituent particles), there is a consensus about few key ``ingredients'' needed to model the evolution of these interesting  objects, i.e. General Relativity, dissipation mechanism and particular equation of state (EoS).  It seems to be useful to consider relatively simple nonstatic models having the above physics constituents, and to analyze some essential features that purely numerical solutions could hinder. This is the rationale behind the present work. Particularly, we will focus on the influence of the dissipation mechanism due to the emission of neutrinos, upon the evolution of compact matter configurations. The effects we study could be a testbed arena for the evolving numerical codes \& simulation tools including General Relativity, dissipation and plausible EoS. 

In the above context, we shall consider the ``thermal peeling'' in general relativistic quasi-static  dissipative anisotropic matter configurations and study under what circumstances it could occur.   This effect, reported several years ago by Herrera and collaborators \cite{HerreraDiPrisco1997}, appears when different signs in the velocity of fluid elements appears, giving rise to the splitting of the matter configuration. These authors shown that, despite it is present in the newtonian regime, it could be fully consistent in the relativistic framework. Recently, thermal peeling has been considered  introducing heat flux in the context called ``separating shells'', which relates the existence of a shell separating an expanding outer region from an inner region that allows collapse towards the center of symmetry \cite{MimosoLeDelliouMena2013,LeDelliouEtal2013}. 

In order to study the conditions that induce the thermal peeling, we shall consider general relativistic spherically symmetric dissipative anisotropic matter distribution in the quasi-static evolution approximation \cite{HerreraDiPrisco1997} and causal second order thermodynamics \cite{Muller1967,Israel1976,Stewart1977}.  
The quasi-static evolution approximation means that changes of the system take place on a time scale that is very long compared to the hydrostatic time scale. As a result of this assumption, physical variables are functions of time but the system may be considered in hydrostatic equilibrium at any moment of its evolution (see \cite{HerreraDiPrisco1997,HerreraEtal1997} and references therein). This approximation is physically reasonable, because most of the lifetime of a compact object could be described on the basis of this quasi-static stages. The hydrostatic equilibrium evolution can be justified in terms of the characteristic times: If the hydrostatic time scale, $\tau _{hyd}\sim  \sqrt{r^3/m}$, is much shorter than the Kelvin-Helmholtz time scale, $\tau_{KH}\sim m^2/2rl$, then inertial terms in the equation of motion $T_{r;\mu}^\mu =0$ can be ignored. This assumption is very sensible because the hydrostatic time scale is remarkably small for almost any phase of a star life. It is of the order of 27 minutes for the Sun, 4.5 seconds for a white dwarf, and $10^{-4}$ seconds for a neutron star of one solar mass and $10 Km$ radius. Therefore, we can be confident that  most relevant processes in star interiors take place on time scales that are usually much longer than the hydrostatic time scale. \cite{HerreraDiPrisco1997,KippenhahnWeigert1994}. More over, in the quasi-static approximation we use, the system is thermally adjusted if it considerably changes its properties only within a characteristic time scale $\tau _{cha}$ that is large as compared with the Kelvin-Helmholtz time scale. In this scenario, the Cattaneo diffusion equation \cite{Cattaneo1948} becomes the approximation of the Israel-Stewart equation \cite{IsraelStewart1979A,IsraelStewart1979B}.

To explore thermal peeling possible scenarios, we carry out a comparative study on the gravitational dissipative collapse for local and quasi-local anisotropic spherical matter configurations, having the same starting static density profile. We call local equation of state (LEoS) the standard barotropic equations of state, $P=P(\rho)$, where the radial pressure is only function of the energy density. On the other side, in a quasi-local equation of state (QLEoS) the radial pressure becomes also function of other state (quasi-local) variables, i.e. $P=P(\rho \, ; \mu, r)$. Quasi-local variables are quantities which values can be derived from the physics accessible within an arbitrarily small but finite neighborhood of a spacetime point and are related to the Misner-Sharp quasi-local mass energy \cite{MisnerSharp1964,CattoenFaberVisser2005,Sussman2008,Szabados2009,HorvatIIijicMarunovic2011B}.  

Thermal peeling were obtained for quasi-static, dissipative mater configurations LEoS and QLEoS, for a very specific energy flux profile with very particular sensivity to the shape of the hight emitted luminosity. If a compact object exhibits this effect it seems to happen when most of the radiated energy comes from the mid layers of the matter distribution and for an wide range of high initial mass of the system. These scenarios are consistent with the one previously reported for a conformally flat quasi-local model \cite{MunozNunez2006}. We did not find any particular sign, in the second order temperature distribution, that could be associated with the peeling effect.

The structure of the present paper is the following. Next section sketches the framework for the quasi-static  approximation describing its consequences for local and quasi-local dissipative anisotropic matter configurations.  In Section \ref{ThermalQLEoSsystems} we shall show the causal second order thermodynamics schema applied.  Following is Section \ref{modellingQLEoSSystems}, where some of the modelling are displayed.  Finally, we close the work with Section \ref{RemarksConclusions} having a brief discussion of the results.

\section{Quasi-static approximation framework}
This section will be devoted to illustrate the concepts involved in the quasi-static  approximation to study the evolution of anisotropic dissipative matter configurations. The influence of local anisotropy (unequal stresses: $P\neq P_{\perp}$) in General Relativity has been extensively studied since the pioneering work of R. Bowers and E. Liang \cite{BowersLiang1974},  (see a classic review on this subject \cite{HerreraSantos1997}). We do not intend to describe the (micro) physical origin for these type of non pascalian fluids. We only consider this type of fluid in order to have the most general picture of a relativistic spherical thermal collapse. 

Following we will present a general relativistic anisotropic dissipative framework and latter we will approximate it to the quasi-static schema.

\subsection{General anisotropic, dissipative spherical framework}
Let us consider an spherically symmetric metric,  
\begin{equation}
\mathrm{d}s^2 = \mathrm{e}^{2 \nu(r,t)}\,\mathrm{d}t^2-\mathrm{e}^{2\lambda(r,t)}\,\mathrm{d}r^2-r^2 \left(\mathrm{d}\theta ^2+
\sin^2\theta\,\mathrm{d}\phi^2\right) \; ,
\label{metricSpherical}
\end{equation}
with the matter described by a heat conducting anisotropic fluid,
\begin{equation}
{T}_{\mu \nu}= (\rho + P_{\perp}){{u}}_\mu{ {u}}_\nu - P_{\perp}{g}_{\mu \nu}  +
(P-P_{\perp}){{v}}_\mu {{v}}_\nu  + {{f}}_\mu{{u}}_\nu +f_\nu {u}_\mu \,,\label{tmunu}
\end{equation}
where 
\begin{equation}
{{u}}_\mu = \gamma (\mathrm{e}^{\nu}, -\omega \mathrm{e}^{\lambda}, 0, 0)  \,, \,
{{v}}_\mu = \gamma (\omega \mathrm{e}^{\nu},-\mathrm{e}^{\lambda}, 0, 0)   \quad \mathrm{and} \quad 
{{f}}_\mu = q\gamma  (\omega \mathrm{e}^{\nu},-\mathrm{e}^{\lambda}, 0, 0)\, . \label{umu}
\end{equation}
Here, $\rho$ is the energy density, $P$ the radial pressure, $P_{\perp}$ the tangential pressure, $q$ is the energy flux, $\omega$ is the radial velocity of the fluid and $\gamma$ is the standard Lorentz factor $\gamma=\left(1-\omega^2 \right)^{-\frac{1}{2}}$. 

The radial proper velocity can be expressed  in this coordinates as:
\begin{equation}
\label{rpuntolocal}
\omega = \frac{\mathrm{d} r}{ \mathrm{d} t} \mathrm{e}^{\lambda -\nu} \; .
\end{equation}

Now, the physical variables, emerging from the Einstein equations with the above metric (\ref{metricSpherical}) and energy momentum tensor (\ref{tmunu}), can be written as:
\begin{eqnarray}
\rho&=& \frac{\mathrm{e}^{-2\lambda} \left[ 2 r \lambda^{\prime}- 1\right]+1 }{8\pi r^{2}\left( 1-\omega^{2}\right)}
+ \frac{ \dot{\lambda}\mathrm{e}^{-\nu-\lambda}  }{2\pi r\left( 1-\omega^{2}\right)} \, \omega 
+\frac{\mathrm{e}^{-2\lambda}\left(2r\nu^{\prime} +1\right)  -1}{8\pi r^{2}\left( 1-\omega^{2}\right)} \omega^{2} \,,
\label{EcCampoGen1}\\
P &=& \frac{\mathrm{e}^{-2\lambda}\left[ 2r \nu^{\prime} +1\right]  -1}{8\pi r^{2}\left( 1-\omega^{2}\right)} 
+\frac{ \dot{\lambda}\mathrm{e}^{-\lambda -\nu}}{2\pi r\left( 1-\omega^{2}\right)}\, \omega
+\frac{\mathrm{e}^{-2\lambda}\left(2r\lambda^{\prime} -1\right)  +1 }{8\pi r^{2}\left( 1-\omega^{2}\right)}\, \omega^{2}
\,,\label{EcCampoGen2} \\
P_{\perp} &=& \frac{e^{-2\lambda}\left(\lambda^{\prime} -\nu^{\prime} 
+r\left(\nu^{\prime} \lambda^{\prime} -\nu^{\prime \prime}  -  \left(\nu^{\prime}\right)^{2}\right)   \right) }{8\pi r} 
+\frac{e^{-2\nu}}{8\pi}\left[ \left(\ddot{\lambda}-\left(\dot{\nu} -\dot{\lambda} \right) \dot{\lambda}\right) \right] \; \; \mathrm{and} \label{EcCampoGen3} \\
q &=& -\frac{ \dot{\lambda}\mathrm{e}^{-\nu-\lambda} }{4\pi r\left( 1-\omega^{2}\right)}  
-\frac{e^{-2\lambda}\left( \lambda + \nu \right)^{\prime}  }{4\pi r\left( 1-\omega^{2}\right)}\omega 
- \frac{\dot{\lambda}\mathrm{e}^{-\nu-\lambda} }{4\pi r\left( 1-\omega^{2}\right)} \omega^{2}
; \label{EcCampoGen4}
\end{eqnarray}
where the dots and primes denote derivatives with respect to $t$ and $r$, respectively.  

If $\nu$ and $\lambda$ are fully specified, then equations (\ref{EcCampoGen1}-\ref{EcCampoGen4}) become a system of algebraic equations for the physical variables $\rho$, $P$, $P_{\perp}$, $\omega$ and $q$. Obviously, in the most general case the system is underdetermined, and two equations of state,  $P=P(\rho)$ and $P=P(P_{\perp})$, should be provided and, in the non-adiabatic case, i.e. $q\neq 0$, a transport equation has to be assumed. For the adiabatic ($q = 0$) and locally isotropic fluid ($P = P_{\perp}$) the system is overdetermined and a constraint on the physical variables appears.

Notice how the right hand side of equations (\ref{EcCampoGen1}),  (\ref{EcCampoGen3}) and (\ref{EcCampoGen4}) are splited in terms of order $\omega$. This separation will be evident in the coming section where second order terms, $O(\omega^{2})$ will be neglected and some restrictions on metric elements.
Next sections will outline the formalism for quasi-static evolving gravitational matter configurations. First, we will consider standard local anisotropic dissipative configurations and next, quasi-local matter distribution (i.e. matter configurations having a QLEoS from now on) will be described.

\subsection{Quasi-static evolution approximation framework}
The simplest situation, when dealing with self-gravitating spheres, is that of equilibrium (static case), i.e. $\omega = q = 0$ and all time derivatives vanishes. The next case is the quasi-static regime, which means that the sphere changes slowly, on a time frame that is very long compared to the typical hydrostatic time scale.
From the above field equations (\ref{EcCampoGen1})-(\ref{EcCampoGen4}) we can obtain
\begin{equation}\label{hydrostatic}
\ddot{\lambda}+{\dot{\lambda}^2}-{\dot{\lambda} \dot{\nu} }=4\pi r{\rm e}^{2\nu}\left[P^{\prime}+{(\rho +P)\nu^{\prime}}-\frac{2(P_\perp -P)}{r}\right]  
\end{equation}
which, under the quasi-static assumptions: $\omega^{2} \approx 0$ and $
\ddot{\lambda}\approx\dot{\lambda}^2\approx\dot{\lambda}\dot{\nu}\approx\ddot{\nu}\approx 0
$, yields
\begin{equation}\label{TOVanisotropic}
P^{\prime}+{(\rho +P)\nu^{\prime}}-\frac{2(P_\perp -P)}{r}= 0 \; .
\end{equation}
Thus, the system  evolves through a series of hydrostatic equilibrium in a time scale that is very long compared to the typical hydrostatic time.  In terms of the behaviour of physical variables of the problem, this quasi-static picture implies that the radial velocity, $\omega$, as well as its radial and time derivatives, have to be considered small. Therefore, the products of these quantities (as well as their second time derivatives) will be neglected. From equation (\ref{EcCampoGen4}), it can be deduced that $q\sim \omega$.

An alternative slow evolving schema could also be attained if we only assume  $\omega^{2} \approx 0$ and no other supposition is made about the time variation of the metric coefficients. In this case we will not necessarily  fulfil (\ref{TOVanisotropic}) during the evolution. We shall denote this case as slow evolution because it is the only assumption we have on matter velocity. This scenario is out of the scope of the present work and will be considered elsewhere.

Changing $\lambda(r,t) \rightarrow m(r,t)$ as
\begin{equation} \label{massdefinition}
{\rm e}^{-2\lambda}=1-\frac{2m(r,t)}{r}\, ,
\end{equation} 
we can re-write the above Einstein system(\ref{EcCampoGen1})-(\ref{EcCampoGen4}) in the quasi-static approximation as:
\begin{eqnarray}
\rho &=& \frac{m'}{4\pi r^2} \; , \label{EcCampoQSL1}\\
 P &=&-\frac{m}{4\pi r^3} +\frac{\nu '}{4\pi r}\left(1 - \frac{2m}{r}\right) \; , \label{EcCampoQSL2}  \\
 P_\perp &=& \frac{1}{8\pi} \left\{ \left( 1 - \frac{2m}{r}\right)\left( \nu^{\prime \prime} + (\nu^{\prime})^{2} + \frac{\nu^{\prime} }{r}\right) 
 + \left( \frac{m^{\prime}}{r} - \frac{m}{r^{2}} \right)  \left( \nu^{\prime} + \frac{1}{r} \right)      \right\} \quad \mathrm{and}\label{EcCampoQSL3}\\
q &=& -\frac{\dot{m}{\rm e}^{ -\nu} }{4\pi r^{2}\sqrt{1 -\frac{2m}{r}} } 
-\left(\left( 1 - \frac{2m}{r}\right)  \nu^{\prime} +\frac{m^{\prime} }{r} -\frac{m}{r^{2}} \right) \frac{ \omega }{4 \pi r} \; . \label{EcCampoQSL4}
\end{eqnarray}
It is interesting to reshape (\ref{EcCampoQSL4}) in terms of two other physical variables (\ref{EcCampoQSL1}) and (\ref{EcCampoQSL2}) as 
\begin{equation}
\label{qrhoP}
q = -\frac{\dot{m}{\rm e}^{ -\nu} }{4\pi r^{2}\sqrt{1 -\frac{2m}{r}} } 
-\left(\rho +P \right) \omega  \; .
\end{equation}

Notice that, in general, the Misner mass definition \cite{HernandezMisner1966} ,
\begin{equation} \label{Misnermass}
m(t,r)=\frac{r^2}{2}R^{3}_{232} \quad \Leftrightarrow \quad m(r,t)=4\pi \int ^r_0 T^0_0r^2\mathrm{d}r ,
\end{equation} 
coincides with the above new variable (\ref{massdefinition}).

It is clear that, the system (\ref{EcCampoQSL1})-(\ref{EcCampoQSL3}) can be solved  using the junctions conditions, providing the mass function $m(r,t)$ (or equivalently density profile $\rho(r,t)$) and giving two barotropic equations of state: $P_\perp=P_\perp(\rho)$ and $P=P(\rho)$.  Then equation (\ref{EcCampoQSL4}), can be solved either by providing, the velocity $\omega(r,t)$ and solving $q(r,t)$ or, given the energy flux $q(r,t)$ profile and obtaining $\omega(r,t)$. 

As we have pointed out before, we are interested to study under what circumstances thermal peeling occurs. Therefore we shall proceed in both ways. First providing a velocity distribution $\omega(r,t)$ which represents thermal peeling obtaining the energy flux profile $q(r,t)$ and secondly the other way round. Given a similar $q(r,t)$ obtained in the previous case and, (hopefully) achieving $\omega(r,t)$ which presents peeling. 

 The modelling we shall work out are going to be locally anisotropic in the pressures, having the same $r$ functionality for the mass distribution $m(r,t)$. The next two sections will show different approaches to solve the above system (\ref{EcCampoQSL1})-(\ref{EcCampoQSL3}). In the next section \ref{LAnisotropic} we shall follow Herrera and collaborators, \cite{CosenzaEtal1981} assuming a particular equation of state $P_\perp= P_\perp(\rho)$ and in \ref{NLAnisotropic}  we shall consider a quasi-local equation of state  \cite{HernandezNunezPercoco1999, HernandezNunez2004}.

\subsection{Quasi-static approximation for local anisotropic configuration}
\label{LAnisotropic}
Following  Herrera and coworkers approach, \cite{CosenzaEtal1981}, we shall assume that the equation of state for radial and tangential pressures is:
\begin{equation}\label{anisotropi_equation}
P_\perp -P=C\zeta(P,r)(\rho +P)r^n \; ,
\end{equation}
where $C$ is a parameter that ``measures'' the local anisotropy of the pressures. Thus, (\ref{TOVanisotropic}) can be written as:  
\begin{equation}\label{presion_equation}
P^{\prime}=-(\rho +P)\nu^{\prime}+2C\zeta(P,r)(\rho +P)r^{n-1} \; .
\end{equation}

For simplicity, these authors assumed  
\begin{equation}
\label{simplicity}
\zeta(P,r)r^{n-1}=\nu^{\prime} \; ,
\end{equation} and,  therefore (\ref{presion_equation}) can be re-written as 
\begin{equation}\label{presion_equationII}
P^{\prime}=-h(\rho +P)\nu^{\prime} \; \Rightarrow \; P^{\prime}=-h\left(\frac{m'}{4\pi r^2} +P\right)\frac{m+4\pi r^3P}{r(r-2m)} \; ,
\end{equation}
with $h = 1-2C$. Clearly $h=1$ recovers an isotropic matter distribution and, from (\ref{TOVanisotropic}) an steeper pressure gradient is obtained if $P_\perp < P$, and a harder EoS is obtained, i.e
\begin{equation}
\label{anisotropyindexhard}
   1 < h <  \infty   \quad  \Rightarrow \quad  P_\perp < P \quad \Rightarrow \; \mathrm{harder} \; \mathrm{EoS} \, ,
\end{equation}
and  
\begin{equation}
\label{anisotropyindexsoft}
 -\infty < h < 1   \quad  \Rightarrow  \quad P_\perp > P  \quad \Rightarrow \; \mathrm{softer} \; \mathrm{EoS} \, . 
\end{equation} 

It is worth to be noticed that the present approach is not pretending to describe the microscopic origin of the anisotropy, but it is clearly taking into account the anisotropic effect on the evolution of the matter configuration. Anisotropy could emerge from various physical scenarios and plays important roles at different scales on several astrophysical objects (see \cite{HerreraSantos1997} references therein and the list of the more of hundred works that cite this interesting review).  The above schema has proven to be useful simulating radiation hydrodynamics scenarios where phase transitions are present \cite{HerreraNunez1989} and very recently is has been used to model colapsing scenarios for anisotropic politropic configurations \cite{HerreraBarreto2013} .

To end this section, we shall define the convective (comoving) time differentiation, $\mathrm{D}(\cdot)/\mathrm{D} t$ which shows how the mass inside a particular shell changes with time. Thus, by using (\ref{EcCampoQSL1}), (\ref{EcCampoQSL4}) and (\ref{rpuntolocal}), we obtain 
\begin{equation}
\label{ConvectiveDerivmlocal}
\frac{\mathrm{D} m}{\mathrm{D} t} \equiv \frac{\partial m}{\partial t} + \frac{\partial m}{\partial r} \frac{\mathrm{d} r}{\mathrm{d} t} =
- 4\pi r^2 \left( P \frac{\mathrm{d} r}{\mathrm{d} t} + q \mathrm{e}^{\nu -\lambda} \right) =
- 4\pi r^2 \mathrm{e}^{\nu -\lambda}\left( P \omega + q  \right) \; .
\end{equation}
Even in this very simple model of quasi-static evolving radiation hydrodynamics, the evolution of the mass at each particular shell, $r=constant$, emerges from a combined effect of the mass transfer and the dissipation field. Now, solving $\omega$ from the above  (\ref{ConvectiveDerivmlocal}) we get
\begin{equation} 
\label{SignOmegalocal}
\omega = -\frac{1}{4\pi r^2 P} \left( \mathrm{e}^{\lambda -\nu}  \frac{\mathrm{D} m}{\mathrm{D} t} + 4\pi r^{2} q \right) \; .
\end{equation} 
In the coming section we will consider quasi-static evolving quasi-local dissipative matter configuration.

\subsection{Quasi-static  approximation for quasi-local matter configuration}
\label{NLAnisotropic}
In this section we shall consider a general relativistic spherical matter configuration, having a QLEoS. As it has been pointed out previously  \cite{HernandezNunezPercoco1999, HernandezNunez2004}, we set a relation between the radial pressure and the energy density as
\begin{equation}
P(r)=\rho(r)-\frac{2}{r^{3}}\int_{0}^{r}\bar{r}^{2}\rho(\bar
{r})\ \mathrm{d}\bar{r}
\,\, \Rightarrow \,\,
P(\rho; \, \mu, r)=\rho-\frac{3\mu}{8\pi r^{2}}  \,, \quad \mathrm{with} \; \; \mu = \frac{2m}{r} \,.
\label{QuasiST}
\end{equation}
The above relation extends concept of a barotropic LEoS $P=P(\rho)$ to a QLEoS of the form $P=P(\rho; \mu, r)$ with $\mu$ (the compactness function) and $r$ as aceptable quasi-local variables\cite{CattoenFaberVisser2005,HorvatIIijicMarunovic2011A,HorvatIIijicMarunovic2011B}. 

Additional physical insight for this particular QLEoS can be gained by considering equation (\ref{QuasiST}) re-written as
\begin{equation}
P_{r}(r)=\rho(r)-\frac{2}{3}\left\langle \rho\right\rangle _{r},
\quad \mathrm{with} \quad \left\langle \rho\right\rangle_{r}=\frac{\int_{0}^{r}4\pi\bar{r}^{2}\rho(\bar{r})\ \mathrm{d}\bar{r}}{\frac{4\pi}{3}r^{3}}\ =\frac{M(r)}{V(r)}\ . \label{averagedens1}%
\end{equation}
Clearly the second righthand term represents an average of the function $\rho(r)$ over
the volume enclosed by the radius $r$. Moreover, equation (\ref{averagedens1})
can be easily rearranged as
\begin{equation}
P_{r}(r)=\frac{1}{3}\rho(r)+\frac{2}{3}\ \left(  \rho(r)-\left\langle\rho(r)\right\rangle \right)  = 
\frac{1}{3} \rho(r)+\frac{2}{3}\ \mathbf{\sigma}_{\rho}\ ,
\label{sigmatmunu}%
\end{equation}
where we have used the concept of statistical standard deviation
$\mathbf{\sigma}_{\rho}$ from the local value of energy density. Furthermore,
we may write:
\begin{equation}
P_{r}(r)=\mathcal{P}(r)+2\mathbf{\sigma}_{\mathcal{P}(r)} \quad\mathrm{where}\left\{
\begin{array}
[c]{l}
\mathcal{P}(r)=\frac{1}{3}\rho(r)\qquad\mathrm{and}\\
\\
\mathbf{\sigma}_{\mathcal{P}(r)}=\left(  \frac{1}{3}\rho(r)-\frac{1}%
{3}\left\langle \rho \right\rangle _{r}\right)  =
\left(  \mathcal{P}(r)-{\bar{\mathcal{P}}}(r)\right)  \ .
\end{array}
\right.  \label{sigma0}%
\end{equation}
Therefore, if at a particular point within the distribution the value of the
density, $\rho(r),$ gets very close to its average $\left\langle
\rho(r)\right\rangle $ the equation of state of the material becomes similar
to the typical radiation dominated environment, $P_{r}(r)\approx
\mathcal{P}(r)\equiv\frac{1}{3}\rho(r).$

For general relativistic compact objects, this type of equation of state was  originally proposed by D. G. Ravenhall and C. J. Pethick in 1994 \cite{RavenhallPethick1994}, as a simplifying assumption to approximate neutron star moment of inertia for a variety of equations of state for nuclear matter.  QLEoSs have proven to be very fruitful describing a variety of relativistic astrophysical scenarios \cite{ArrautBaticNowakowski2009, Nicolini2009, HernandezNunez2013}.

The metric functions corresponding to a spherical  matter configuration having a QLEoS are related as $\nu(r,t) = \lambda(r,t) +K(t)$  \cite{HernandezNunezPercoco1999, HernandezNunez2004} and, with this premise we can translate most of the above described quasi-static local evolution  framework.  Thus, we can re-state the above field equations for the quasi-static approximation as:
\begin{eqnarray}
\rho &=& \frac{m'}{4\pi r^2} \; , \label{EcCampoQSNL1} \\
P &=& - \frac{m}{2\pi r^3}+ \frac{m^{\prime}}{4 \pi r^2} \; , \label{EcCampoQSNL2} \\
 P_{\perp} &=&\frac{1}{8 \pi}\left\{\frac{m^{\prime \prime}}{r}+\frac{2(m^{\prime} r-m)}{r^3}\left[\frac{m^{\prime} r-m}{r-2m}-1\right]\right\}
\; \; \mathrm{and}  \label{EcCampoQSNL3} \\
q &=&  -\frac{\dot{m}{\rm e}^{ -\nu} }{4\pi r^{2}\sqrt{1 -\frac{2m}{r}} } 
- \frac{ \left(r\,m^{\prime} -m \right)\omega }{2 \pi r^{3}} \; ; \label{EcCampoQSNL4} 
\end{eqnarray}
with the proper velocity of the fluid which now expressed as: 
\begin{equation}
\omega = \frac{dr}{dt} \, e^{-K} \;.  \label{rpunto}
\end{equation}
Finally, equation (\ref{SignOmegalocal}) can be re-written for the quasi-local case as
\begin{equation}
\label{SignOmega}
\omega = -\frac{1}{4\pi r^2 P} \left( \mathrm{e}^{-K}  \frac{\mathrm{D} m}{\mathrm{D} t} + 4\pi r^2 q \right) \; ,
\end{equation}
preserving the above mentioned description for the radiation hydrodynamics picture.

\subsection{Junction Conditions and the surface evolution}
Before proceeding to describe the causal second order thermodynamic schema for  quasi-static matter configurations, it is interesting to point out some general consequences emerging from the junctions conditions which are valid for both, local and quasi-local matter distributions.

Because we are describing radiating configurations, our inner solutions have to be matched to the exterior Vaidya line element outside the source, i.e. 
\begin{equation}
\label{VaidyaExterior}
\mathrm{d}s^{2}=\left( 1-\frac{2\mathcal{M}(u)}{\mathcal{R}} \right) \mathrm{d}u^{2}+2\mathrm{d}u \;\mathrm{d}\mathcal{R}-\mathcal{R}^{2}\left(\mathrm{d}\theta^2+ \sin^2\theta\,\mathrm{d}\phi^2\right) \,, 
\end{equation}
where $u$ is a timelike coordinate such that $u=const$ is, asymptotically, a null cone open to the future and $\mathcal{R}$ is a null coordinate (i.e. $g_{\mathcal{RR}}=0$). 

Following \cite{HerreraDiPrisco1997} we matched the quasi-static  interior solution to the exterior Vaidya radiating metric (\ref{VaidyaExterior}) at $r = \mathcal{R}=R$, thus
\begin{eqnarray}
m(R) & = & M \label{masscontinue}\\
{\rm e}^{ \nu_{R}}  & = & 1- \frac{2M}{R} \label{nucontinue}\\
\left[ P \right]_{R}  & = &\left[ q \right]_{R}  \label{Pcontinue}
\end{eqnarray}
where the subscript $R$ denotes that the variable is evaluated at the surface of the distribution, and $\left[ var \right]_{R}$ represents the discontinuity of the variable across $R$. 

Now evaluating the last field equation (\ref{qrhoP})  at $r=R(t)$, we obtain
\begin{equation}\label{EvalEEsurface}
\left.\frac{\partial\, m(r,t)}{\partial t}\right|_{r=R(t)}=-4\pi R(t)^2 \left(1-\frac{2M(t)}{R(t)}\right) \left[ \omega_R (\rho_R + P_R) +q_R \right]  \; .
\end{equation}
Next, we expand $m(r,t)$ and evaluate it at the surface, 
\begin{equation}
\label{mExpand}
\left.\frac{\partial m(r,t)}{\partial t}\right|_{r=R(t)}=\dot{M}(t)-\left.m^\prime\right|_{r=R(t)} \; \dot{R}(t)  \; ,
\end{equation}
and by substituting  equations (\ref{EcCampoQSL1}), (\ref{EcCampoQSL2}) and (\ref{mExpand}) into (\ref{EvalEEsurface}), we finally get,
\begin{equation}\label{dotM}
\frac{\mathrm{d}M(t)}{\mathrm{d}t}=- 4\pi R(t)^2 \left(1-\frac{2M(t)}{R(t)}\right) q_R = -L\; .
\end{equation}
This equation relates the total luminosity, $L$, as measured at infinite with the evolution of the boundary, the surface gravitational potential and the energy flux evaluated at the surface of the configuration. It is clear that from (\ref{dotM}) we can solve the evolution of the boundary of the distribution as a function of the luminosity and the total mass, i.e.
\begin{equation}
\label{SufEvolution}
R(t) = M(t) \left( 1 \pm \sqrt{ 1 - \frac{\dot{M}}{4 \pi \, q_{R} \; M^{2}} } \right) \quad \mathrm{with} \; \dot{M} \equiv \frac{\mathrm{d}M(t)}{\mathrm{d}t} \; ,
\end{equation}
which seems to set bounds on the functional form of the luminosity profile. But for radiating compact objects, normally we have $\dot{M} < 0$ and $q_{R} > 0$ therefore no restriction is imposed by equation (\ref{SufEvolution}). This is a general results independent of the particular equation of state depending upon the assumption of the quasi-static evolution.

Next section will be devoted to describe the second order thermodynamic schema and how the quasi-static  approximation is applied to it. 
%
%

\section{Second order thermal conduction}
\label{ThermalQLEoSsystems}
Concerning the causal second order thermodynamic, it is well known that the Maxwell-Fourier law leads to a parabolic equation (diffusion equation) which presents pathologies on the propagation of perturbations (see \cite{JosephPreziosi1989,JouCasas-VazquezLebon1988,Maartens1996,Muller1999,HerreraPavon2002} and references therein). To overcome such difficulties, various relativistic theories with non-vanishing relaxation times have been proposed in
the past. Since the pioneering work by M\"uller  \cite{Muller1967}, followed by the covariant formulation of W. Israel and J.M. Stewart \cite{Israel1976,Stewart1977,IsraelStewart1979A,IsraelStewart1979B}, all these theories provide a heat transport equation which is not of Maxwell-Fourier type, instead it is based on hyperbolic equation for the propagation of thermal perturbations. 

In this section we shall describe the causal second order schema describing the propagation of thermal perturbations and we will show how it could be implemented for the quasi static evolution framework. The corresponding hyperbolic transport equation for the heat flux, in M\"uller-Israel-Stewart theories reads \cite{IsraelStewart1979A,IsraelStewart1979B}
\begin{equation}
\label{IsraelStewart}
\tau \frac{Df^\alpha}{Ds}+f^\alpha =\kappa P^{\alpha\beta}(T_{,\beta}-Ta_{\beta}) -\tau u^{\alpha}f_{\beta}a^{\beta}-\frac{1}{2}\kappa T^2\left(\frac{\tau}{\kappa T^2}u^{\beta}\right)_{;\beta}f^{\alpha} \; ,
\end{equation}
where $P^{\alpha\beta}$ is the projector onto the three space orthogonal to fourvelocity $u^{\beta}$; $f^{\alpha}$ is the energy flux quadric-vector,  $a^{\beta}$ the quadri-aceleration and $T(r,t)$ the temperature distribution within the matter configuration. Here, there are two important parameters that will be considered in details below, i.e.  $\kappa$ which denotes the (neutrino) thermal conductivity  and $\tau$ describing the relaxation time. 
Due to the symmetry of the problem, equation (\ref{IsraelStewart}) has only one independent component, i.e. 
\begin{eqnarray}
  \label{IsraelStewartComp}
& &q{\rm e}^{\lambda}\left(1+\frac{\tau}{2}u^{\beta}_{;\beta}\right)(1-\omega^2)^{1/2}
+\tau \left({\rm e}^{\lambda-\nu}\dot{q} +\omega q^{\prime}\right) 
+  \kappa (\omega\dot{T} {\rm e}^{\lambda-\nu} +T^{\prime})= \\ \\
 & & \qquad \qquad - \kappa  T\left[{\rm e}^{\lambda-\nu}\left(\frac{\dot{\omega}}{1-\omega^2}+\omega \dot{\lambda}\right)
 +\nu^{\prime} +\frac{\omega \omega^{\prime}}{1-\omega^2} \right] 
 +\frac{q\tau}{T}\left( {\rm e}^{\nu-\lambda}\dot{T}+\omega T^{\prime} \right) \; , \nonumber
 \end{eqnarray} 
   with
\begin{equation}
u^{\alpha}_{\ ; \alpha} = 
\frac{{\rm e}^{-\lambda}}{\left({1- \omega^{2}}\right)^{\frac12} } \left[ {{\rm e}^{\lambda-\nu }} 
\left( {\frac {\omega \dot{\omega} }{1-\omega^{2}}}+ \dot{\lambda}  \right) +\omega\nu' +{\frac { \omega' }{1-\omega^{2}}}+
{\frac {2\omega  }{r}} \right] \,.
\label{Du}
\end{equation}

Following the Herrera and collaborators approach \cite{HerreraDiPrisco1997,HerreraEtal1997,HerreraMartinez1998},  system is always in a hydrostatic equilibrium or very close to it and its evolution may be regarded as a sequence of static models.  In these consecutive quasi-static configurations the fluid is thermally adjusted, i.e. it is in {\it complete equilibrium} as it is stated in ref \cite[p.66]{KippenhahnWeigert1994}. 

Again, the radial velocity, as seen by a minkowskian observer is assumed small, which means that quadratic and higher terms in $\omega$ may be neglected in a first order perturbation theory. It is clear that this condition will be accomplished for small values of luminosity and consequently also for small values radiation flux within the configuration.
In this quasi-static evolution approximation Israel-Stewart transport equation (\ref{IsraelStewartComp}) becomes: 
\begin{equation}\label{SlowCatt2}
q(r,t)=-\kappa\left(T'(r,t)+T(r,t)\nu^{\prime} \right){\rm e}^{-\lambda} \; ,
\end{equation}
which leads to 
\begin{equation}\label{SlowCatt2NL}
q(r,t)=-\kappa\left(T'(r,t)+T(r,t)\lambda '\right){\rm e}^{-\lambda} \; ,
\end{equation}
for the quasi-local case. 

In order to integrate (\ref{SlowCatt2}) or (\ref{SlowCatt2NL}) we need to specify the thermal conductivity $\kappa$ which is also function of the temperature.  Following  \cite{Weinberg1971,Martinez1996}, we shall assume 
\begin{equation}
\label{thermalconductivity}
\kappa = \frac{4}{3} b\, T^{3}\, \tau_{c} \, ,
\end{equation}
where $\tau_{c}$ is the radiation mean collision time (neutrinos in our case) measured by a comoving observer and $b$ a constant which, for neutrinos, can be written as $b = \frac{7\, \mathbf{s}}{8}$, with $\mathbf{s}$ the usual Stefan-Boltzmann constant \cite{Weinberg1971}. 

Again, following reference \cite{Martinez1996}, we are assuming that:  
\begin{itemize}
  \item the propagation speed of thermal signals are of the order of sound speed, 
  \item the neutrinos are generated by thermal emission ($\epsilon_{\nu} \approx  \mathbf{k}_{B}T$ )  and
  \item the radiation mean collision time for neutrinos can be written as
\begin{equation}
\label{neutrinocomvmeantime}
\tau_{c} = \frac{\mathcal{A} M_{0}}{[\rho]\sqrt{\mathcal{Y}_{e} T^{3}}  } \; ,
\end{equation}
with $\mathcal{A} \approx 10^{9} \; \mathrm{K}^{3/2} \; \mathrm{m}^{-1}$; $M_{0}$ the initial mass in meters; $\mathcal{Y}_{e}$ is the electron fraction; $T$ the temperature in Kelvin and $[\rho]$ the dimensionless energy density. More over, the mean collision time for an observer moving with a velocity $\omega$, can be expressed as
\begin{equation}
\label{neutrinocomvmeantime}
\tau = \sqrt{1-\omega^{2}} \, \tau_{c} =   \sqrt{1-\omega^{2}} \, \frac{ \mathcal{A} M_{0}}{[\rho]\sqrt{\mathcal{Y}_{e} T^{3}}  } \; .
\end{equation} 
\end{itemize}
Thus, based on the above considerations, we may also assume a power-law behavior for the thermal conductivity as
\begin{equation}
\label{kappapowerlaw}
\kappa \approx 2.98 \times 10^{-38} \frac{\sqrt{1-\omega^{2}}}{[\rho]} T^{\xi} \; .
\end{equation}
Now, from  (\ref{SlowCatt2}) and (\ref{SlowCatt2NL}) it can be formally integrated yielding  the corresponding  temperature profile, $T(r, t)$, as:
\begin{equation}
\label{ThermalProfile}
T(r,t)= {\rm e}^{-\nu(r,t)} \left( \int_{R(t)}^{r} \frac{(4-\xi) [\rho(\tilde{r},t)]q(\tilde{r},t){\rm e}^{\lambda(\tilde{r},t) +(1+\xi) \nu(\tilde{r},t)}  }{2.98 \times 10^{-38}  \sqrt{1-\omega(\tilde{r},t)^{2}} } \, \mathrm{d}\tilde{r} + u(t)\right)^{\frac{1}{1+\xi}} 
\end{equation}
and 
\begin{equation}
\label{ThermalProfileNL}
T(r,t)= {\rm e}^{-\lambda(r,t)} \left( \int_{R(t)}^{r} \frac{(1-\xi)[\rho(\tilde{r},t)] q(\tilde{r},t){\rm e}^{(1+\xi) \lambda(\tilde{r},t)}  }{2.98 \times 10^{-38} \sqrt{1-\omega(\tilde{r},t)^{2}} } \, \mathrm{d}\tilde{r} + \tilde{u}(t)\right)^{\frac{1}{1+\xi}} \; ,
\end{equation}
respectively. In both of the above relations  (\ref{ThermalProfile}) and  (\ref{ThermalProfileNL}), we shall use $\xi = \frac{3}{2}$ which is the case for neutrino thermal conductivity \cite{Martinez1996}.  The integration functions  $ u(t)$ and $ \tilde{u}(t)$ can be found through the boundary conditions at the surface of the distribution. 

It is clear that luminosity $L$ as measured at infinite is related with the emitted energy as
\begin{equation}
\label{LuminosityEnergy}
L = -\dot{M}(t) \quad \Rightarrow E = L(1 -z_{0})^{2} = L \left( 1 - \frac{2 M(t)}{R(r)}\right)^{-1} \, ,
\end{equation} 
with $z_{0}$ the gravitational redshift.  
By recalling the Stephan-Boltzmann law, it is obtained that the emitted energy is
\begin{equation}
\label{StephanBoltzmannLaw}
E=4\pi \; \mathbf{s} \;R(t)^2T^4_R  \;  
\end{equation}
and by using (\ref{dotM}) we get the temperature at the boundary of the matter distribution as
\begin{equation}
\label{SurfaceTemp}
T_{R}^{4}(t) = \frac{q(R(t),t)}{\mathbf{s} } \equiv \frac{-\dot{M}(t) }{4\pi \; \mathbf{s} \;R(t)^2}  \left( 1 - \frac{2 M(t)}{R(r)}\right)^{-1} \, .
\end{equation}

\section{Modelling quasi-static thermal evolution}
\label{modellingQLEoSSystems}
Our modelling will focus on quasi-static dissipative matter distributions, with local and  quasi-local equation of state, having density profiles with the same dependence in $r$.
The dynamics and the thermal quasi-static evolution of the configuration will be studied by using a particular heat flux function that generates the thermal peeling. The radiating matter configurations are matched, at the surface, with two emission luminosity profiles. Finally, we will discuss the contribution of the QLEoS to the thermal quasi-static evolution, as comparing it with the evolution of other local anisotropic configurations having equivalent density profiles. 

It should be stressed that,  it is possible to obtain all the physical variables from both of the above Einstein systems, i.e. 
 (\ref{EcCampoQSL1})-(\ref{EcCampoQSL4}) for LEoS and  (\ref{EcCampoQSNL1})-(\ref{EcCampoQSNL4}) for QLEoS, once the energy density profile, $\rho(r,t)$, and the energy flux, $q(r,t)$ are given, In the case of local anisotropic models, the pressure, $P(r)$, should be integrated from equation (\ref{presion_equationII}) and then, $ P_\perp$ and $\omega$ can be obtained provided the energy flux profile. For the quasi-local Einstein System, clearly  $\rho(r,t)$, and $q(r,t)$ are the functions to be provided.

Despite we are going to maintain the same $r$-dependence for the local and quasi-local modelling, they might slightly differ in the time functions which can be  solved from the boundary conditions. As in any physical situation, several dynamics models could emerge from the same $r$-dependences of the seed equation of state\cite{AguirreHernandezNunez1994}. 

The next two sections will be devoted to discuss some of the possible energy density and radiation energy flux profiles. 

\subsection{Density profiles}
\label{DensityProfiles}
Now follows the different density profiles to model quasi-local and local quasi-static evolving dissipative matter configurations. We shall workout models emerging from two particular density profiles inspired by the static solutions of Florides \cite{Florides1974} (rediscovered by Gokhroo \& Mehra \cite{GokhrooMehra1994}, FGM-solution from now on) and by H.B. Buchdahl \cite{Buchdahl1959}. The physical variables for the quasi-local physical variables will be denoted as $\rho^{*}$, $P^{*}$, $\omega^{*}$ and $q^{*}$.

\subsubsection{Florides-Gokhroo-Mehra quasi-local density profile}
The FGM-solution under particular circumstances \cite{Martinez1996}, gives rise to the  Bethe-B\"{o}rner-Sato equation state for nuclear matter
\cite{Demianski1985, ShapiroTeukolsky1983, BetheBornerSato1970}. This solution is also a particular  form of the VII Tolman family of solutions \cite{Tolman1939}. 
We shall generalize this density profile as
\begin{equation}
\rho(r)= {\bar{\rho}}_c\left[1- \frac{\mathcal{K}}{R^2} \; r^{2}\right] \quad  \longrightarrow \quad 
\rho^{*}_{F}(t,r)= \rho^{*}_{Fc} \left[1- f^{*}_{F} r^{2}\right] \, ,
\label{Fdin}
\end{equation}
with $f^{*}_{F}= f^{*}_{F}(t)$ and $\rho^{*}_{Fc}(t) = \rho^{*}_{Fc}(t) $. 

From the above generalization we can get,  via equations (\ref{EcCampoQSNL1}) and (\ref{EcCampoQSNL2}), the expressions for the mass $m(r, t)$ and the radial pressure $P (r, t) $ as:
\begin{equation}
m^{*}_{F}(r,t)=4\pi r^3\rho^{*}_{Fc}\left[\frac{1}{3}-\frac{r^2 f^{*}_{F}}{5} \right] \quad \mathrm{and} \quad P^{*}_{F}(r,t)=\frac{\rho^{*}_{Fc}}{3}\left[1-\frac{3r^2f^{*}_{F}}{5}\right] \; ,
\end{equation} 
with the velocity distribution from (\ref{EcCampoQSNL4}) written as
\begin{equation}
\omega^{*}_{F} =  \frac{3  \left( \left(  \left( 10 - 6 r^2 f^{*}_{F} \right) r \dot{\rho}_{c{F}} -6 r^3  \dot{f}^{*}_{F} \; \rho^{*}_{Fc}    \right){\rm e}^{-K} - \frac{45q}{6 \pi r^2 }\right) }
{4 \rho^{*}_{Fc}  \left( 25 -27f^{*}_{F} r^{2} \right)}
\end{equation}

\subsubsection{Buchdahl quasi-local density profile}
This solution was discovered by H.B. Buchdahl \cite{Buchdahl1959} and
rediscovered later by Durgapal \& Bannerji in 1983 \cite{DurgapalBannerji1983}. The corresponding static seed density
profile, which is free from singularities at the origin, can be written and generalised as:
\begin{equation}
\label{rhoBuchdalhlNL}
\rho(r)=\frac{3C}{16\pi}\,{\frac{3+C{r}^{2}}{\left(  1+C{r}^{2}\right)  ^{2}}} \quad  \longrightarrow \quad
 \rho^{*}_{B}(r,t) =\frac{ \rho^{*}_{Bc}(t)}{3}\frac{3+f^{*}_{B}r^2}{\left(1+f^{*}_{B}r^2\right)^2} \; .
\end{equation}
Again,  $f^{*}_{B}= f^{*}_{B}(t)$ and $ \rho^{*}_{Bc} =  \rho^{*}_{Bc}(t)$.  We can also get expressions for the mass $m(r, t)$ the radial pressure $P (r, t) $ as:  
\begin{equation}
\label{mPressBuchdalhlNL}
m^{*}_{B}(r,t)= \frac{4 \pi}{3}\left( \frac{ \rho^{*}_{Bc}(t) r^{3}}{3+f^{*}_{B}r^2}\right)  \quad  \mathrm{and}  \quad 
P^{*}_{B}(r,t)= \frac{ \rho^{*}_{Bc}(t)}{3} \left(\frac{1-f^{*}_{B}r^2}{\left(1+f^{*}_{B}r^2\right)^2}\right)  \; .
\end{equation}
Notice that at the center of the distribution we get, $\rho^{*}_{Bc} = 3 P^{*}_{Bc}$, i.e. a pure radiation equation of state. Once more, from  (\ref{EcCampoQSNL4}), the velocity distribution can be written as:
\begin{equation}
\omega^{*}_{B} = \frac{1}{4 {\rm e}^{K} \rho^{*}_{Bc} }
\left( 
\rho^{*}_{Bc} \dot{f}^{*}_{B}\, r^{3}  - \dot{\rho}^{*}_{Bc}\,r  \left(1+f^{*}_{B}\,r^2\right) 
-3q{\rm e}^{K}\left(1 + 2 f^{*}_{B}\,r^2 +\left( f^{*}_{B}\right)^{2}r^{4} \right)
 \right). 
\end{equation}

\subsubsection{Florides-Gokhroo-Mehra local anisotropic density profile}
Starting, again from the static density profile of the FGM-solution static profile, we get 
\begin{equation}
\label{FLocalMod}
\rho(r)=\rho_{Fc}\left(1-\frac{5r^2}{9R^2}\right) \;  \rightarrow 
\rho_{Fc}=\rho_{Fc}(t)\left(1-\frac{5r^2}{9R(t)^2}\right) \;
\end{equation} and
\begin{equation}
m_{F}(r,t)=\frac{4\pi r^3 \rho_{F}(t)}{9}\left(3-\frac{r^2}{R(t)^2}\right) \quad \mathrm{with} \quad
\rho_{Fc}(t)=\frac{9M(t)}{8\pi R(t)^3} \; ,
\end{equation}
but the pressure, $P(r)$, should be numerically integrated from equation (\ref{presion_equationII}). Then $ P_\perp$ and $\omega$ can be obtained provided the energy flux profile.

\subsubsection{Buchdahl local anisotropic density profile}
Finally, let us consider again the Buchdahl static density profile, then we can get 
\begin{equation}
\label{BuchdahlLocalMod}
\rho(r)=\frac{3C}{16\pi}\,{\frac{3+C{r}^{2}}{\left(  1+C{r}^{2}\right)  ^{2}}} \;  \longrightarrow \;
 \rho_{B}(r,t) =\frac{ \rho_{Bc}(t)}{3}\frac{3+\frac{16 \pi \rho_{Bc}(t)}{9}r^2}{\left(1+\frac{16 \pi \rho_{Bc}(t)}{9}r^2\right)^2}
\end{equation}
and
\begin{equation}
m_{B}(r,t) = \frac{12 \pi r^{3}  \rho_{Bc}(t) }{9 +16 \pi \rho_{Bc}(t) r^{2}} \quad \mathrm{with} \quad
\rho_{Bc}(t)= \frac{9M(t)}{4 \pi R^{2}(t) \left( 4M(t) - 3 R(t) \right) } \; ,
\end{equation}
as in the previous case  the pressure, $P(r)$, has to be numerically integrated from equation (\ref{presion_equationII}) and $ P_\perp$ and $\omega$, again can be obtained provided the energy flux profile. Notice that here, in (\ref{BuchdahlLocalMod})  the generalization of the static density profile involves only one function of $t$, while in the quasi-local case (\ref{rhoBuchdalhlNL}) we need two to match density and pressure to the boundary conditions.

\subsection{Energy flux and luminosity profiles}
\label{fluxprofiles}
For the present paper we shall explore the thermal and dynamic effects coming out from a particular energy flux function having the following profile: 
\begin{equation}
\label{heatflowPeeling}
q= \frac{1}{M_{0}^{2}} \left\{{\frac {79}{11492205}}\, \left( \frac{r}{M_{0}} +{\frac {33035}{26572}} \right) \frac{r}{M_{0}} 
 \left( \frac{r}{M_{0}} -{\frac {53056}{53047}} \right)  \left( \frac{r}{M_{0}} -{\frac {51508}{
48797}} \right)  \left( 1-{\frac {3\, \tau}{100000}} \right) \right\}
\\
\end{equation}
 with $M_{0} = M(t=0)$.
 
\begin{figure}[h]
   \begin{center}
\subfigure[Energy Flux Profile]   
{\includegraphics[scale=0.32]{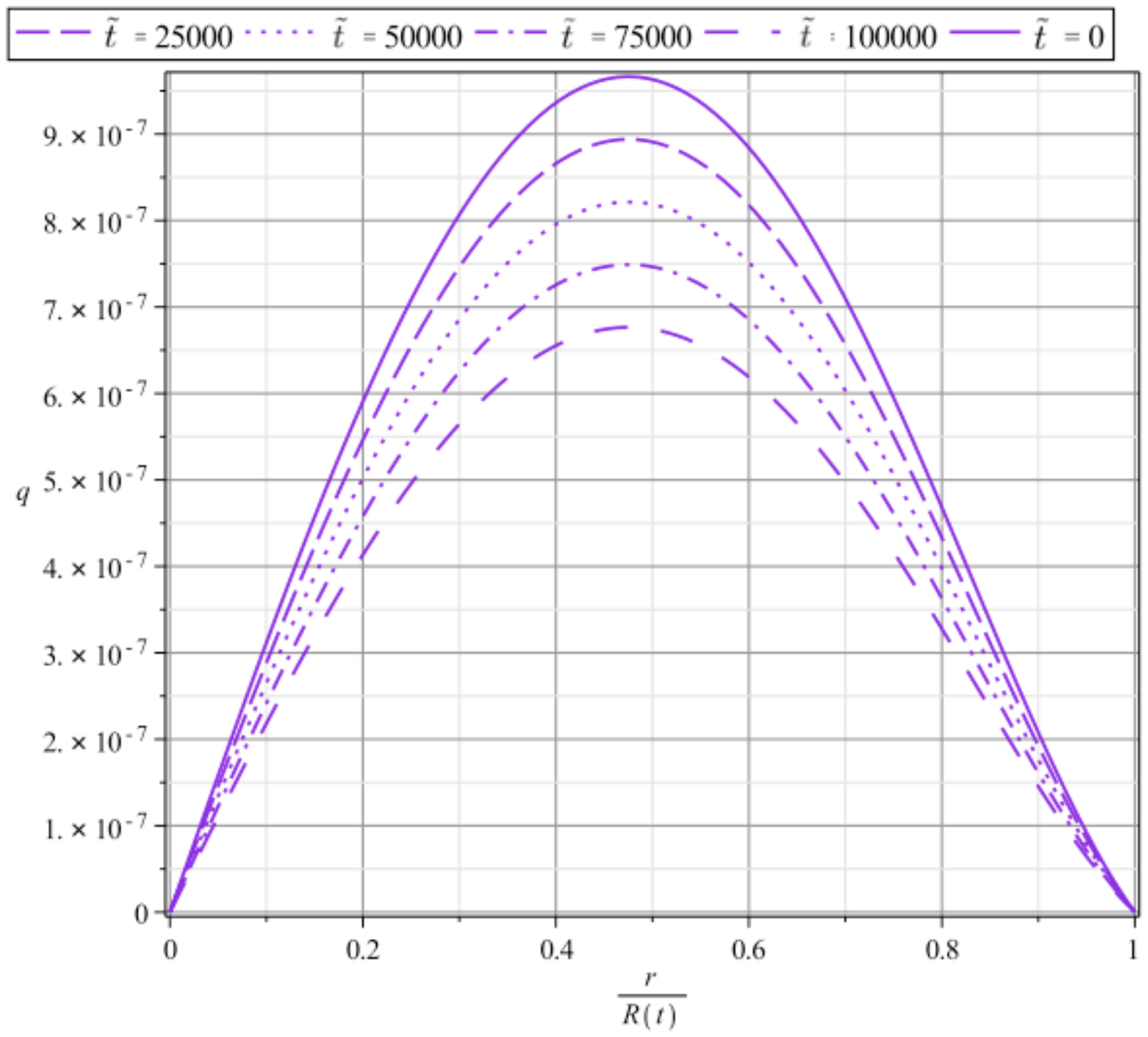} \label{V05Fig1FluxProfile}} \\ 
\subfigure[Luminosity Profiles]
{\includegraphics[scale=0.32]{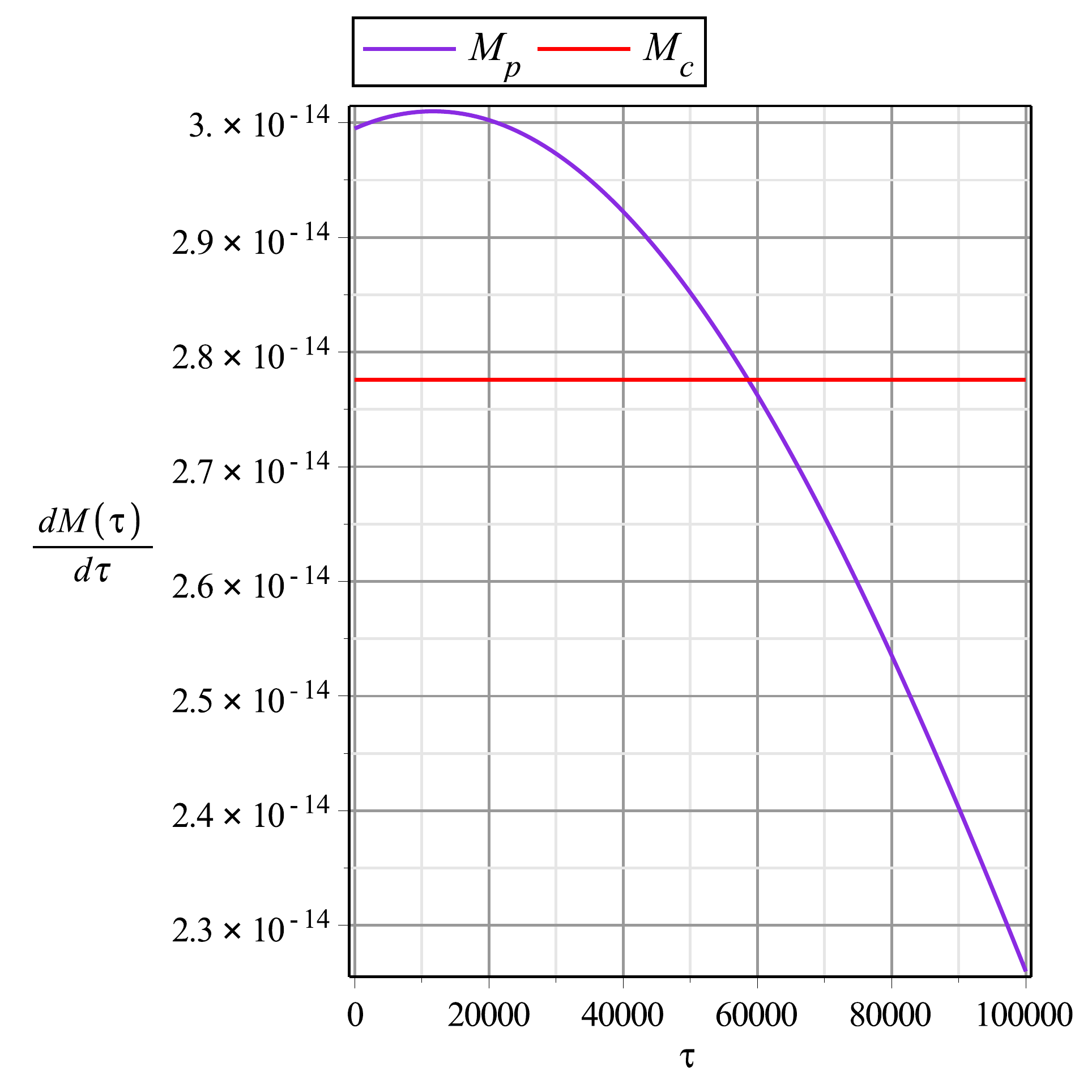} \label{LuminosityProfilesGrf}} \\
\caption{The luminosity profiles are displayed in plate \ref{LuminosityProfilesGrf}, i.e. constant and a pulse-decreasing shape. We have set the same ejected mass for both profiles and the selected pulse parameters for equation (\ref{PulseProfile2}) are $\sigma = 3.5\times 10^{5}$, and $\tilde{t}_{peak} = 3.5 \times 10^{4}$. In plate \ref{V05Fig1FluxProfile} the energy flux (Eq. (\ref{heatflowPeeling})) profiles for different $\tilde{t}$ are sketched. It corresponds to particular flux function, having a maximum at the mid layers of the matter distribution and responsible of a thermal peeling for quasi-static evolving distributions considered.} 
  \end{center}
  \end{figure}

This particular energy flux function, displayed in Figure \ref{V05Fig1FluxProfile}, generates the thermal peeling shown in all quasi-static evolving configurations considered.   As it was mentioned previously, we have implemented the above profile and obtained the peeling effect but we have also worked it out in other complementary way.  We have also provided a velocity distribution having a peeling effect and obtained a heat flux with a similar profile shown in Fig. \ref{V05Fig1FluxProfile}. Thus,  we can consider that this shape for the energy flux profile can be associated with the thermal peeling.
 
The present work will also assume two particular functions for the luminosity  at the surface of the distribution which are displayed in Figure \ref{LuminosityProfilesGrf}. First, we assume a constant luminosity, $L_{c}$ profile and, secondly a pulse-like shape for the radiation emission, $L_{p}$, i.e.  
\begin{equation}
\label{PulseProfile1}
L_{c} \equiv \frac{\mathrm{d}\tilde{M}(\tilde{t})}{\mathrm{d}\tilde{t}}= -\frac{\tilde{M}_{e}}{\tilde{t}_{f}}
\end{equation} 
and
\begin{equation}
\label{PulseProfile2}
L_{p} \equiv \frac{\mathrm{d}\tilde{M}(\tilde{t})}{\mathrm{d}\tilde{t}}=-\left( \frac{8 \pi \tilde{M}_{e}}{\sigma M_{0} \sqrt{2\pi} } \right)  {\rm e}^{-\frac{1}{2}\left(\frac{\tilde{t}-\tilde{t}_{peak}}{\sigma}\right)^2},
\end{equation} 
respectively. 

Notice that $\tilde{M}$, $R$ and $\tilde{t}$ are dimensionless quantities  because we have scaled them by the initial mass,   $M_{0}$, i.e.: $M(\tilde{t})=\tilde{M}(\tilde{t})\times M_{0} $, $R(\tilde{t})= \tilde{R}(\tilde{t}) \times M_{0} $ and  $t = \tilde{t} \times M_{0}$. The other physical parameters we use, are:  
\begin{itemize}
  \item $\tilde{t}_{f}$ is the cutoff luminosity time,
  \item $\tilde{t}_{peak}$ is the time for the maximum for the emission luminosity pulse,
  \item $\tilde{M}_{e}$ as the total amount of ejected mass through the luminosity ($\tilde{M}(\tilde{t}_{f})-\tilde{M}(\tilde{t}=0)$) and  
  \item $\sigma$ is the standard deviation of the gaussian radiation pulse.
\end{itemize}
Clearly, equation (\ref{SurfaceTemp}) can be written as
\begin{equation}
\label{SBunits}
L \equiv \frac{\mathrm{d}\tilde{M}(\tilde{t})}{\mathrm{d}\tilde{t}}=-4\pi \; \mathbf{s} \; M^{2}_{0} \; \tilde{R}(\tilde{t})^2 \; T^4_R 
\left( 1 - \frac{2 \tilde{M}(\tilde{t})}{\tilde{R}(\tilde{t})}\right) \; ,
\end{equation} 
where $T_R$ is the surface temperature of the configuration. 

\subsection{The modelling}
Our modelling will be focused on the evolution of the radial velocity profile, its relation with the energy flux distribution and temperature profile within the matter distribution. Table \ref{SimulParam}, presents the set of physical parameter we have chosen for our modelling. L-B and QL-B stand for local and quasi-local Buchdahl models, while L-FGM and QL-FGM represent local and quasi-local Florides-Gokhroo-Mehra configurations, respectively.

\begin{table}[h!]
  \centering 
\begin{tabular}{|l|l|l|l||l|l|l|}
\hline
 Simulation Parameters   & \multicolumn{2}{c|}{L-B} & QL-B &  \multicolumn{2}{c|}{L-FGM} & QL-FGM \\  
 \cline{2-3}  \cline{5-6} 
 $ \left(1 -2M_{0}/R_{0}\right) \approx 0.23 $ & $h= 1.1$ & $h= 1/80$ &  & $h= 1.1$  & $h= 1/80$ &\\ \hline  \hline
$M_{0} = 2M_{\odot}; \, R_{0}=7.8Km$  &  &  &  &  &  &\\ \hline 
$  \rho_{c} \times 10^{15} \; g/cm^{3} $ & $\approx 4.16$   &  $\approx 4.14$    & $\approx 4.06$  
		& $\approx 3.04$  & $\approx 3.04$  & $\approx 3.04$  \\ \hline
$  P_{c} \times 10^{34} \; din/cm^{2} $ & $ \approx 364.27$  & $ \approx 1.36 $ & $ \approx 121.74$ 
		& $ \approx 546.13 $ & $ \approx 1.16 $ & $ \approx 91.31 $ \\ \hline
 $M_{e}  \times 10^{-9} \; M_{\odot}$  & $\approx 1.99$  & $\approx 1.99$ & $\approx 1.99$ & $\approx 1.99$ & $\approx 1.99$ & $\approx 1.99$ \\ \hline 
 $T_{R}(0) \times 10^{9} K$  & $\approx 8.57 $  & $\approx 8.57$  & $\approx 8.57$  & $\approx 8.57$ & $\approx 8.57$  &$\approx 8.57$  \\ \hline  \hline
\hline
$M_{0} = 20M_{\odot}; \, R_{0}=78Km$  &  &  &  &  &  &\\ \hline 
$  \rho_{c} \times 10^{13} \; g/cm^{3} $ & $\approx 4.16$   &  $\approx 4.16$    & $\approx 4.06$  
		& $\approx 3.04$  & $\approx 3.04$  & $\approx 3.04$  \\ \hline
$  P_{c} \times 10^{21} \; din/cm^{2} $ & $ \approx 4985.50$  & $ \approx 9.09 $ & $ \approx 812.03$ 
		& $ \approx 3642.7 $ & $ \approx 7.67 $ & $ \approx 609.02 $ \\ \hline
 $M_{e}  \times 10^{-9} \; M_{\odot}$  & $\approx 1.99$  & $\approx 1.99$ & $\approx 1.99$ & $\approx 1.99$ & $\approx 1.99$ & $\approx 1.99$ \\ \hline 
 $T_{R}(0) \times 10^{8} K$  & $\approx 8.57 $  & $\approx 8.57$  & $\approx 8.57$  & $\approx 8.57$ & $\approx 8.57$  &$\approx 8.57$  \\ \hline  \hline
\hline
\end{tabular}   
\caption{This table presents the set of physical parameter we have chosen for our modelling. L-B and QL-B stand for local and quasi-local Buchdahl models, while L-FGM and QL-FGM represent local and quasi-local Florides-Gokhroo-Mehra configurations, respectively.  We have denoted: $\rho_{c}$ as the central density; $P_{c}$ for the central pressure; $M_{e}$ as the total amount of ejected mass; $M_{0}, R_{0}$ and $T_{R}(t=0)$ are used for the initial mass, initial radius and initial surface temperature of the matter configuration. Finally, $M_{\odot}$ indicates the solar mass and $h$ represents the ``degree'' of anisotropy of local pressures, having $h=1$ for isotropy.}
\label{SimulParam}
\end{table}
The luminosity profiles are displayed in plate \ref{LuminosityProfilesGrf}, i.e. constant and a pulse-decreasing shape. We have set the same ejected mass for both profiles and the selected pulse parameters for equations (\ref{PulseProfile2}) are $\sigma = 3.5\times 10^{5}$, and $\tau_{peak} = 3.5 \times 10^{4}$. The total dimensionless time, $\tau = t/M_{0}$, sets the order of magnitude of typical evolution time of our simulations to 
$0 \lesssim t \lesssim 1s$. 

The energy flux distribution distribution within the matter configuration is sketched in 
Fig \ref{V05Fig1FluxProfile}. It shows that most of the radiated energy comes from the mid layers of the distribution.

\begin{figure}[h!]
\centering
{\includegraphics[scale=0.31]{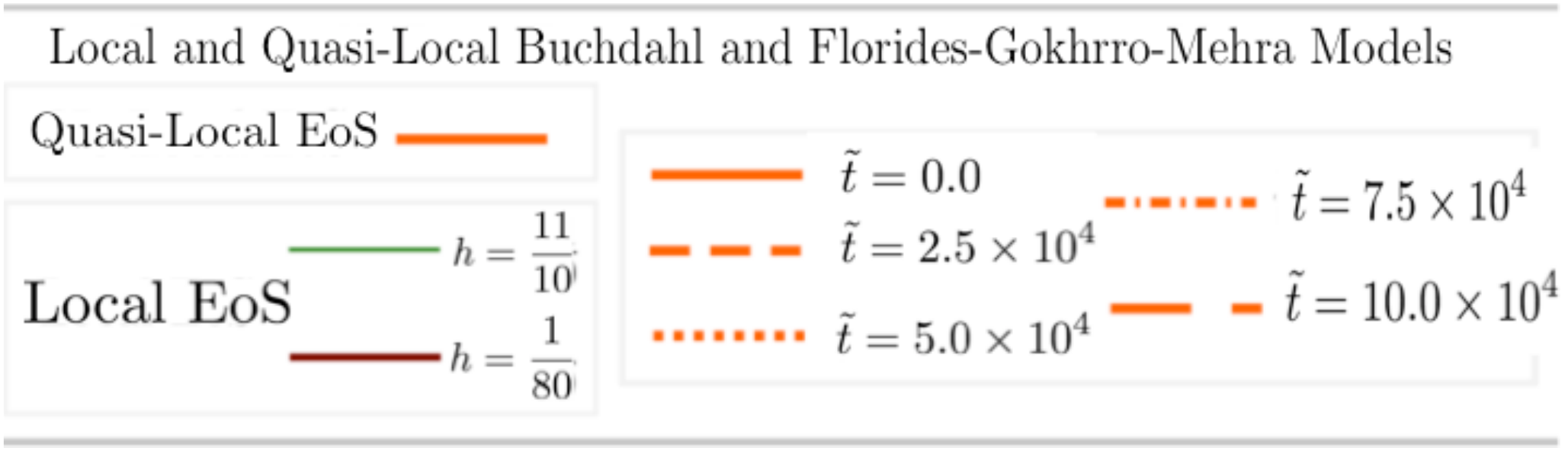}}  \\
\subfigure[B$\, L_{c}$]
{\includegraphics[scale=0.22]{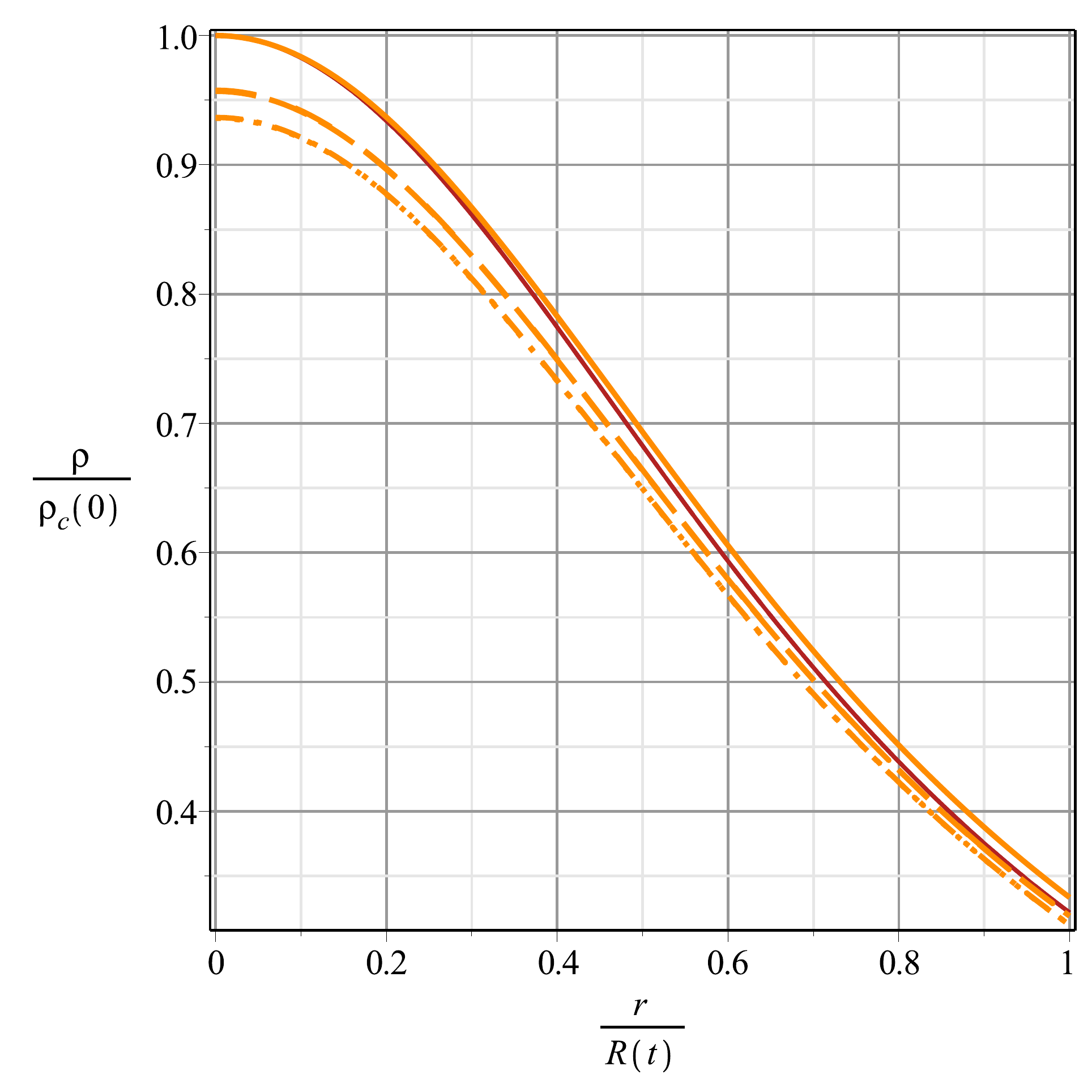} \label{rhoBc} }
\subfigure[F$\, L_{c}$]
{\includegraphics[scale=0.22]{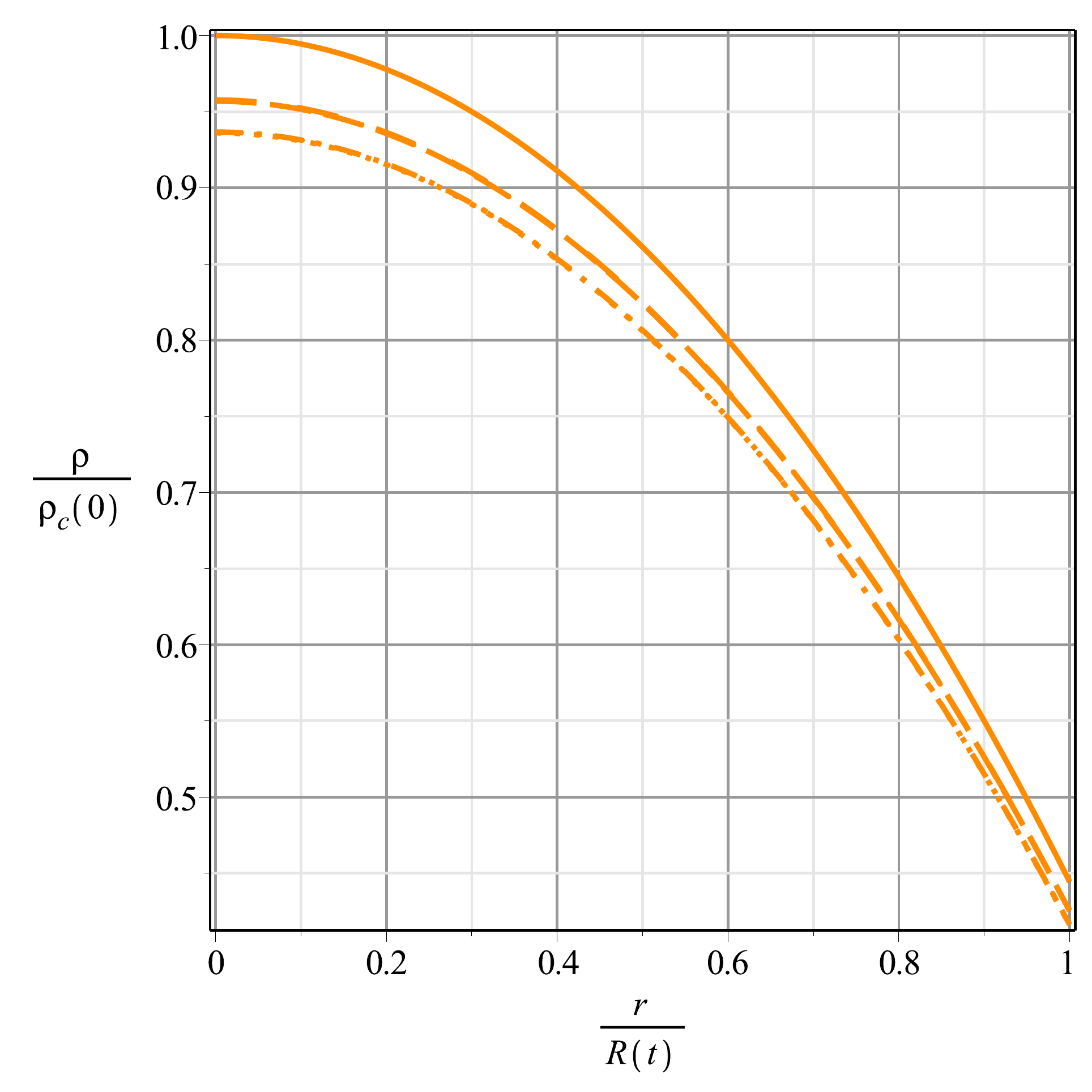} \label{rhoFc} }

\subfigure[B$\, L_{p}$]
{\includegraphics[scale=0.22]{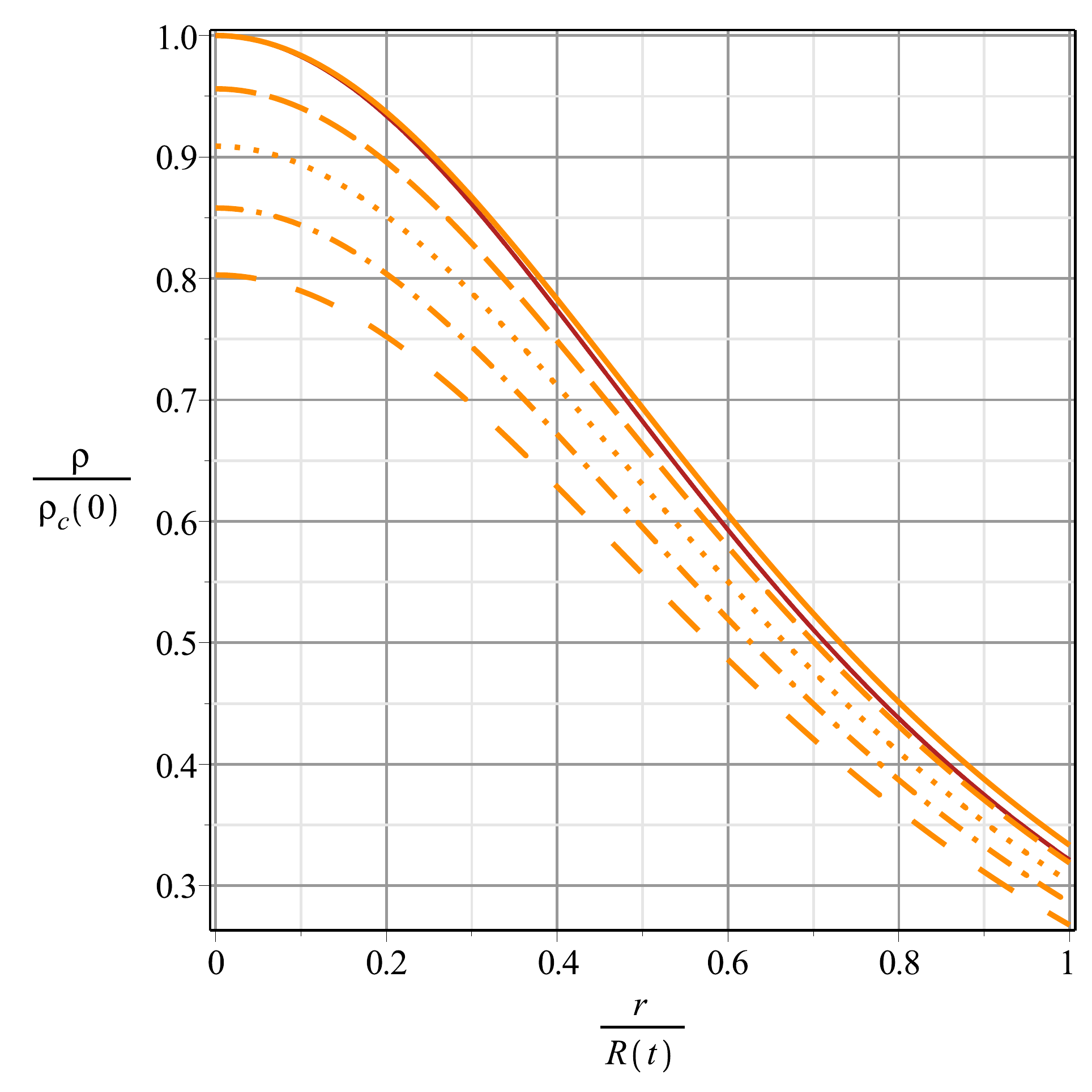} \label{rhoBp} }
\subfigure[F$\, L_{p}$]
{\includegraphics[scale=0.22]{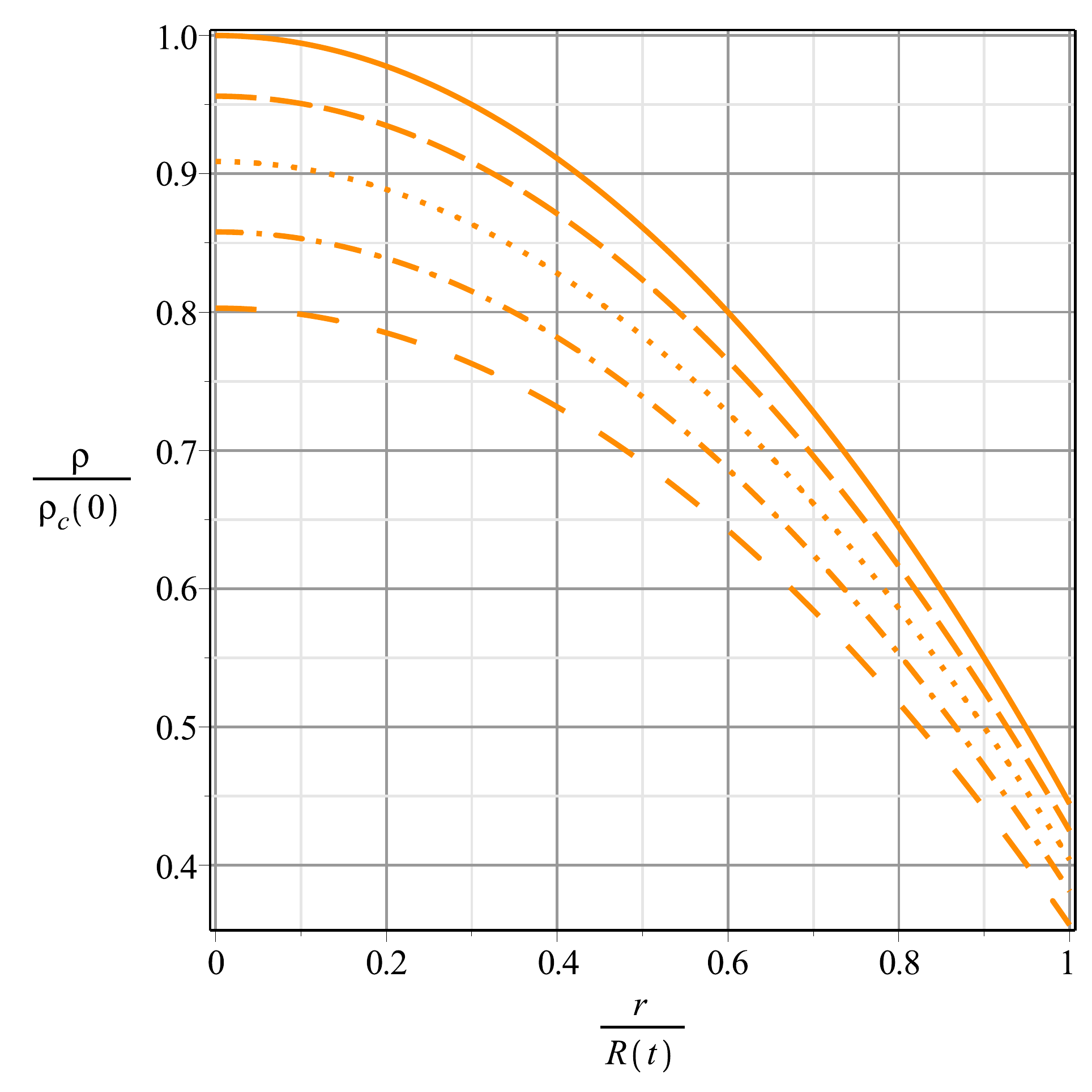}\label{rhoFp} }
\caption{Profiles for physical variable, $\rho(r,t)$ scaled by a any central initial density $\rho_{c}(t=0)$. Plates \ref{rhoBc} and \ref{rhoFc} stand for constant luminosity, while \ref{rhoBp} and \ref{rhoFp} represent those with a pulse-like one. All density profiles displayed are almost the same for LEoS and QLEoS 
}
\end{figure}

\begin{figure}[h!]
\centering
\includegraphics[scale=0.31]{V60CaptionAll.pdf}  \\
\subfigure[B$\, L_{c}$]
{\includegraphics[scale=0.22]{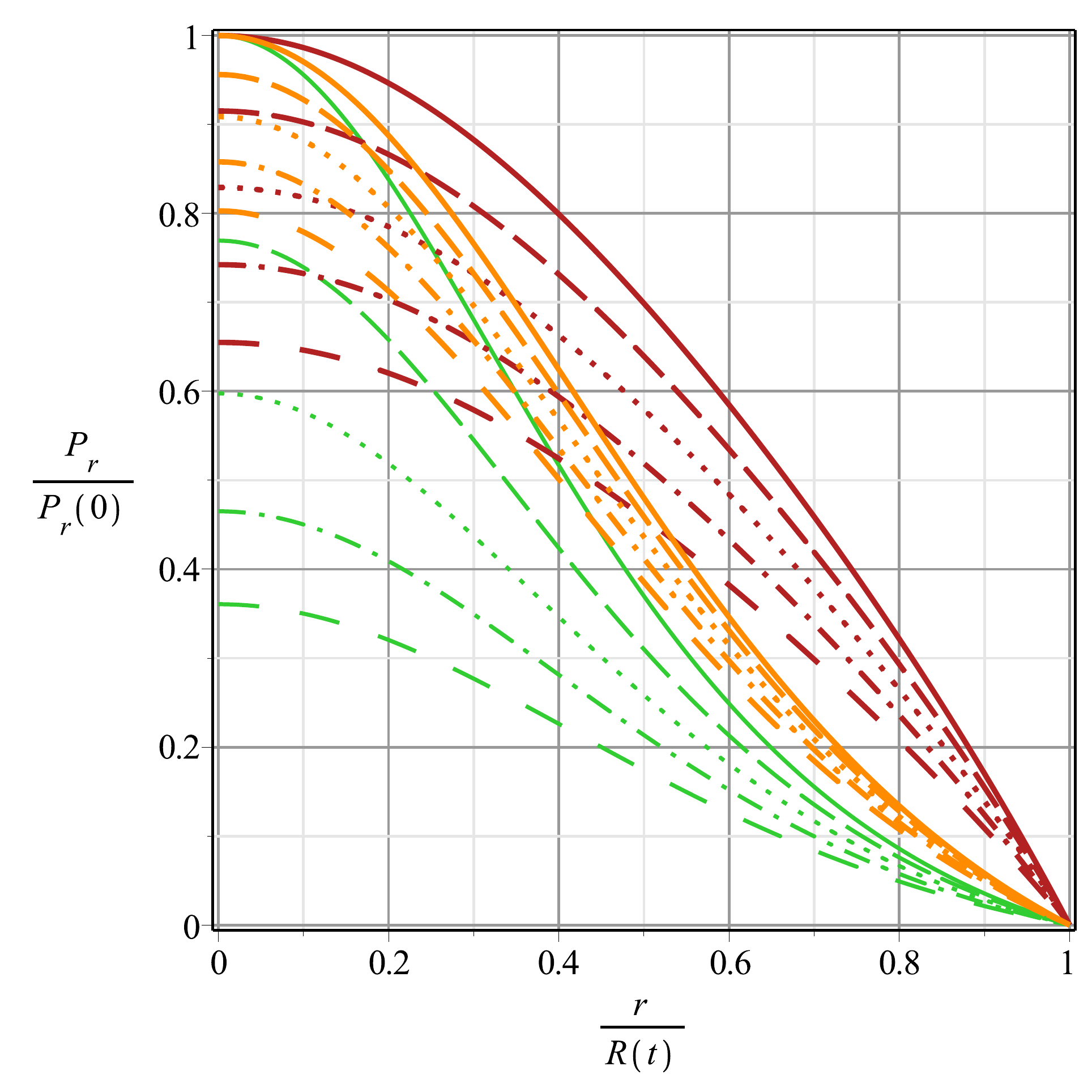} \label{PBc} }
\subfigure[F$\, L_{c}$]
{\includegraphics[scale=0.22]{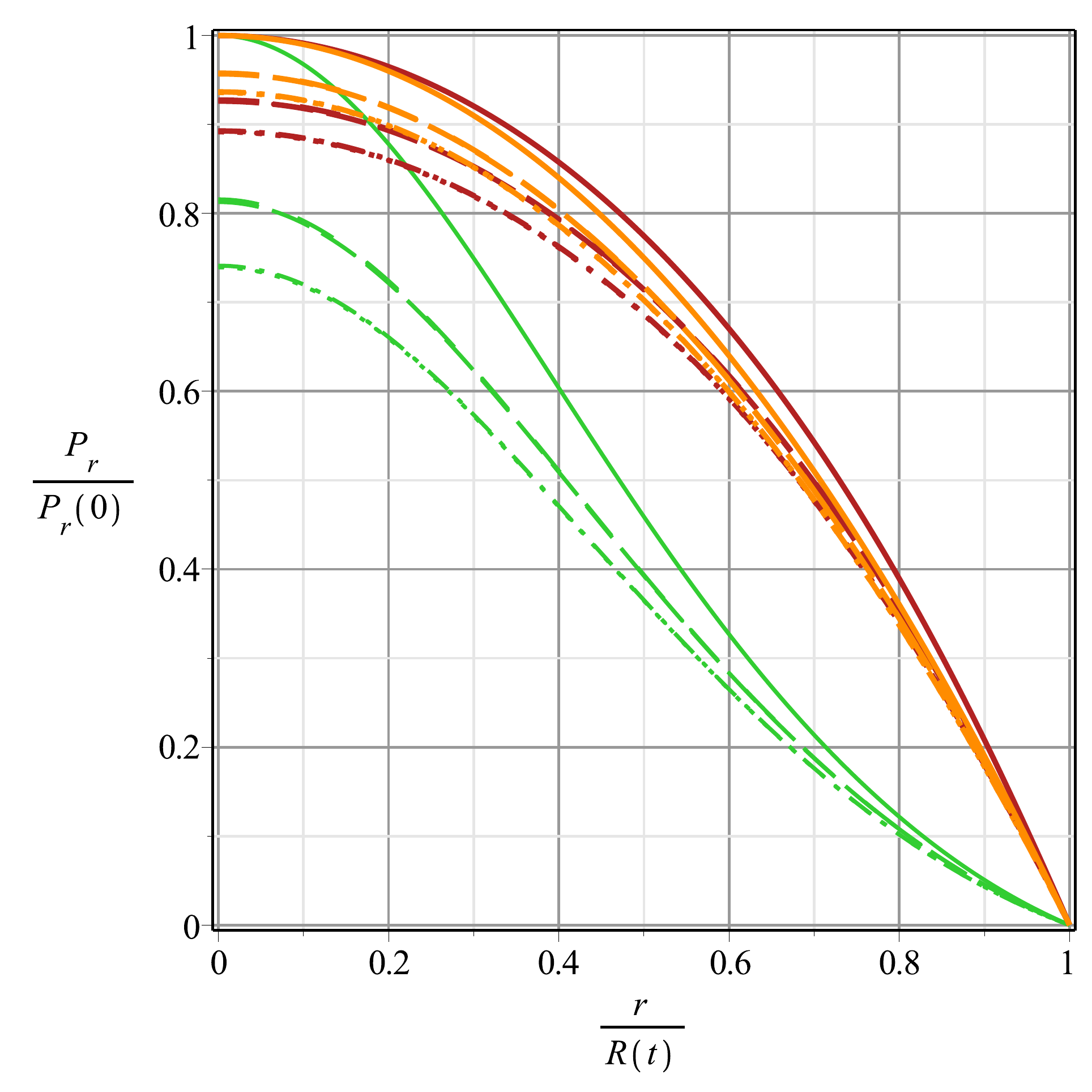}\label{PFc} } 

\subfigure[B$\, L_{p}$]
{\includegraphics[scale=0.22]{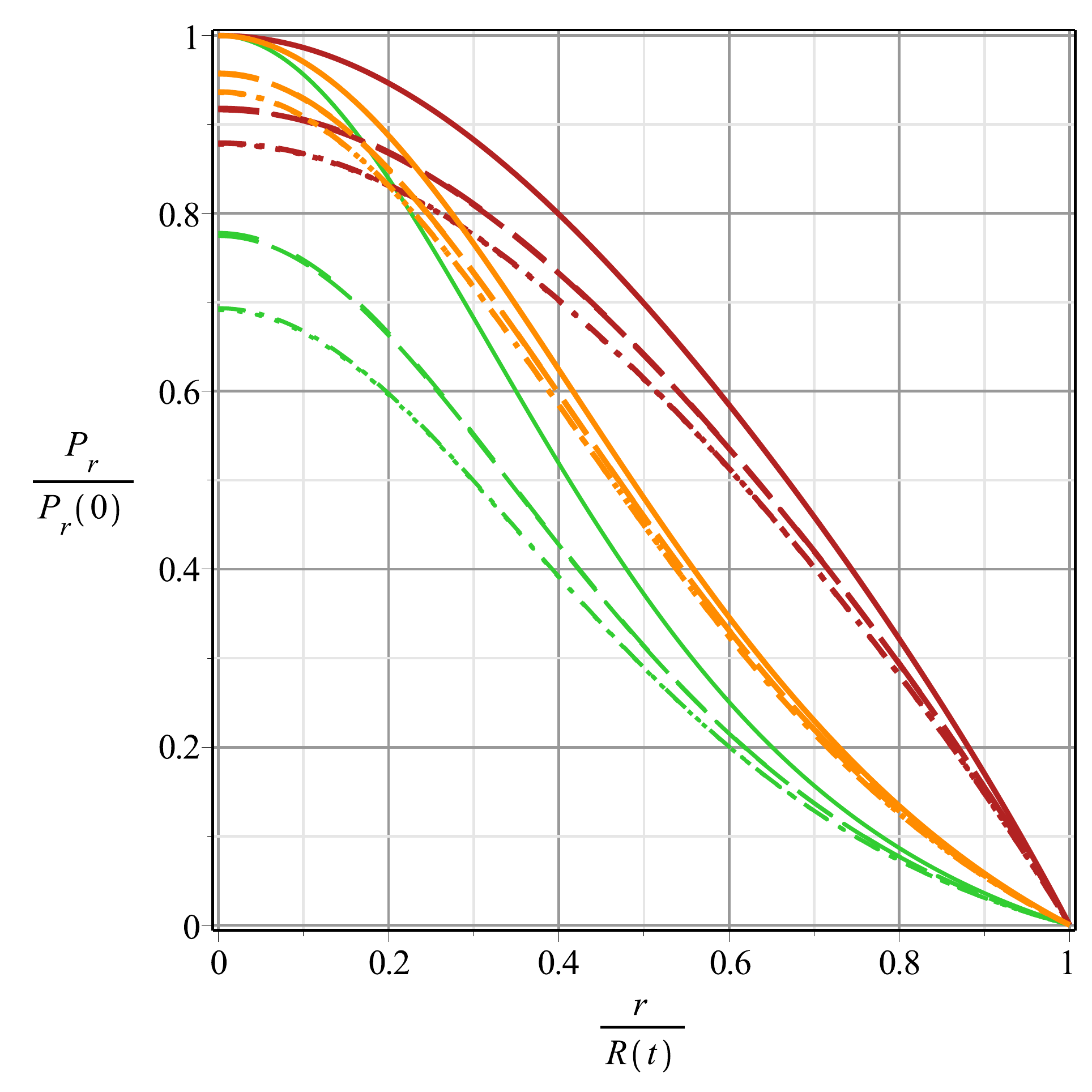}\label{PBp} }
\subfigure[F$\, L_{p}$]
{\includegraphics[scale=0.22]{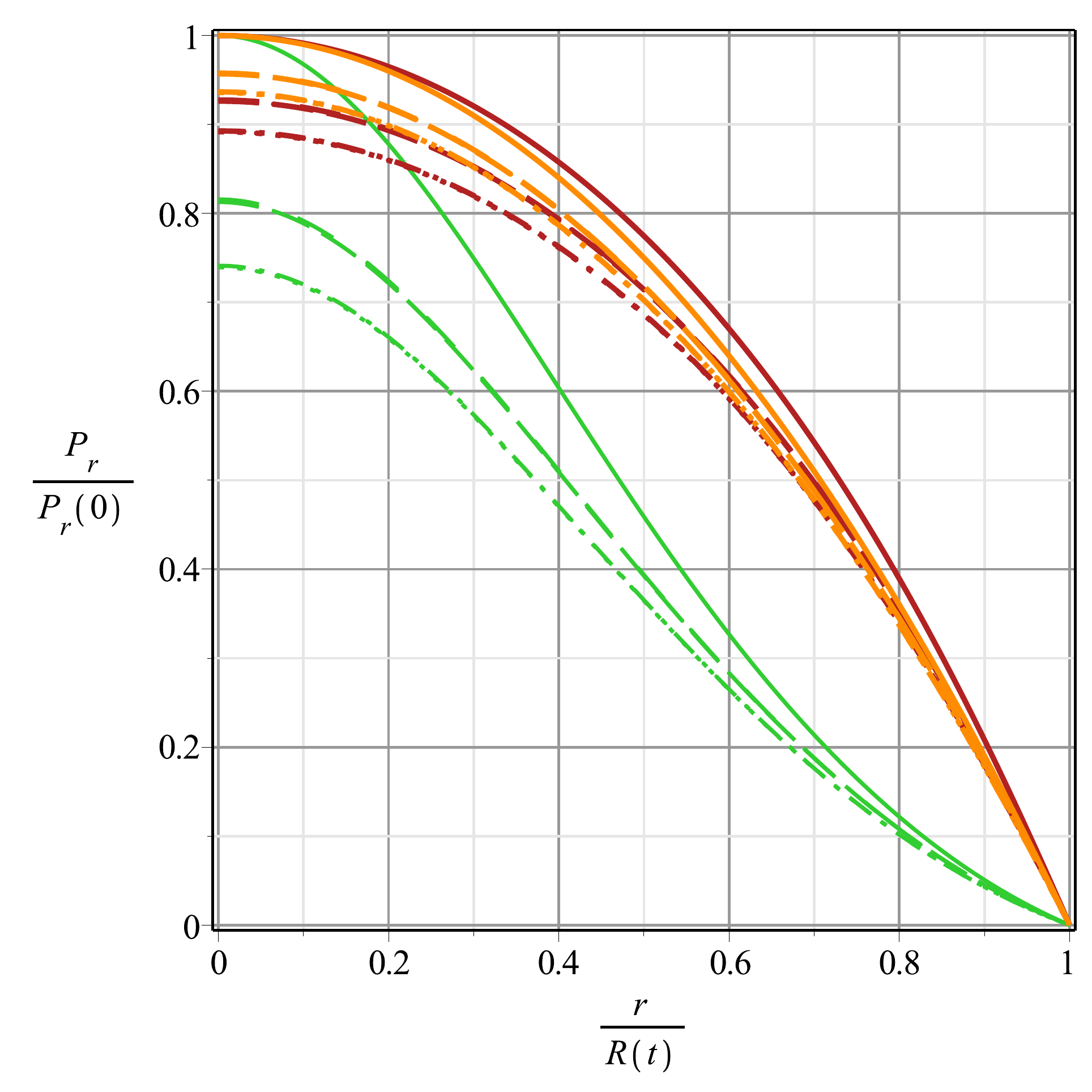}\label{PFp} }
\caption{Profiles for the radial pressure $P(r,t)$ scaled by the central initial pressure $P_{c}(t=0)$. Plates \ref{PBc} and \ref{PFc} stand for constant luminosity, while \ref{PBp} and \ref{PFp} represent those with a pulse-like one. It is clear the distinction between a \textit{hard} ($h=11/10$) and \textit{soft} ($h=1/80$) LEoS. The pressure gradient, $\mathrm{d} P(r,t)/\mathrm{d}r$, for QLEoS models are in the middle between the \textit{hard} and the \textit{soft} anisotropic LEoS. } 
\end{figure}

\begin{figure}[h!]
\centering
\includegraphics[scale=0.31]{V60CaptionAll.pdf}  \\
\subfigure[B$\, L_{c}$]
{\includegraphics[scale=0.22]{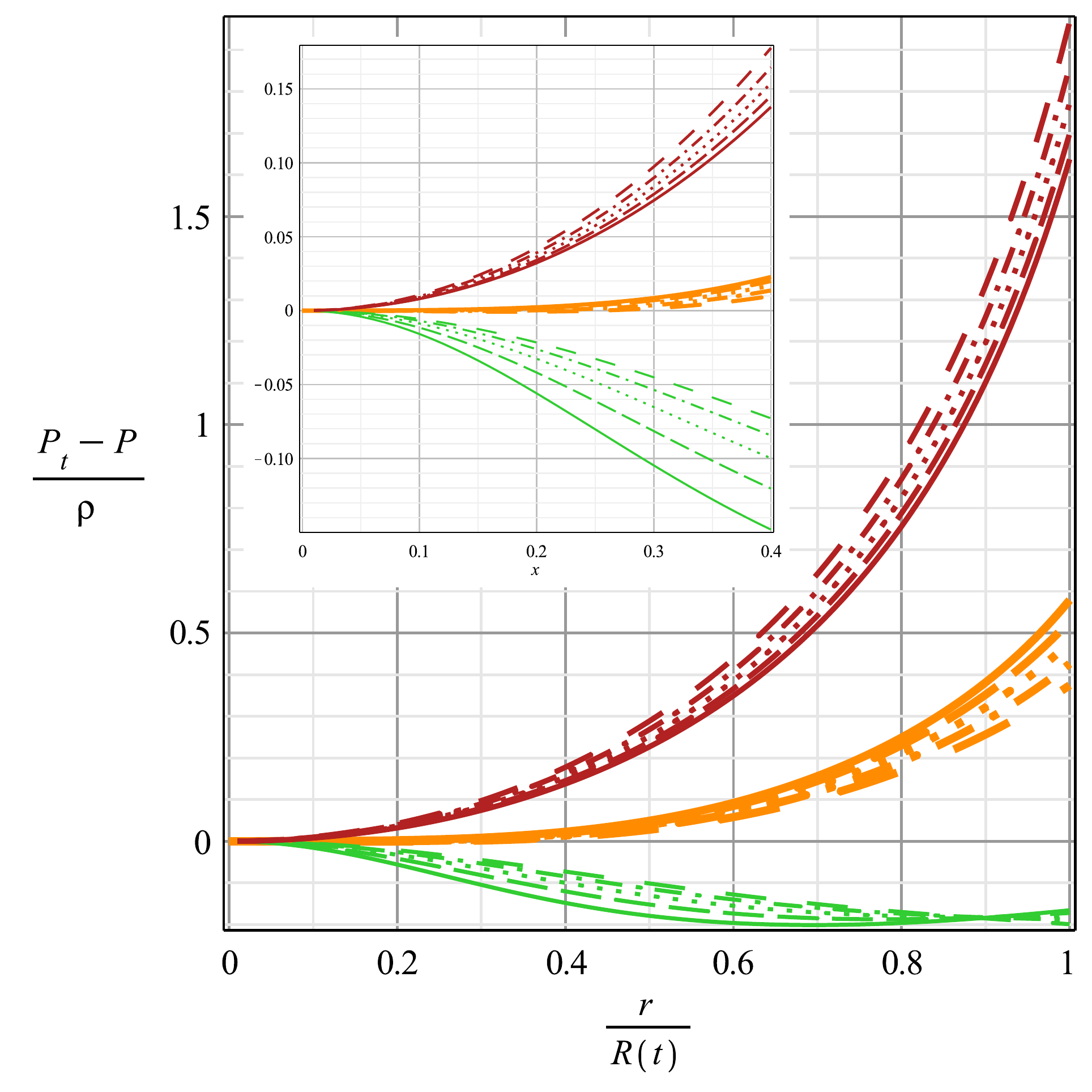}\label{PtBc} }
\subfigure[F$\, L_{c}$]
{\includegraphics[scale=0.22]{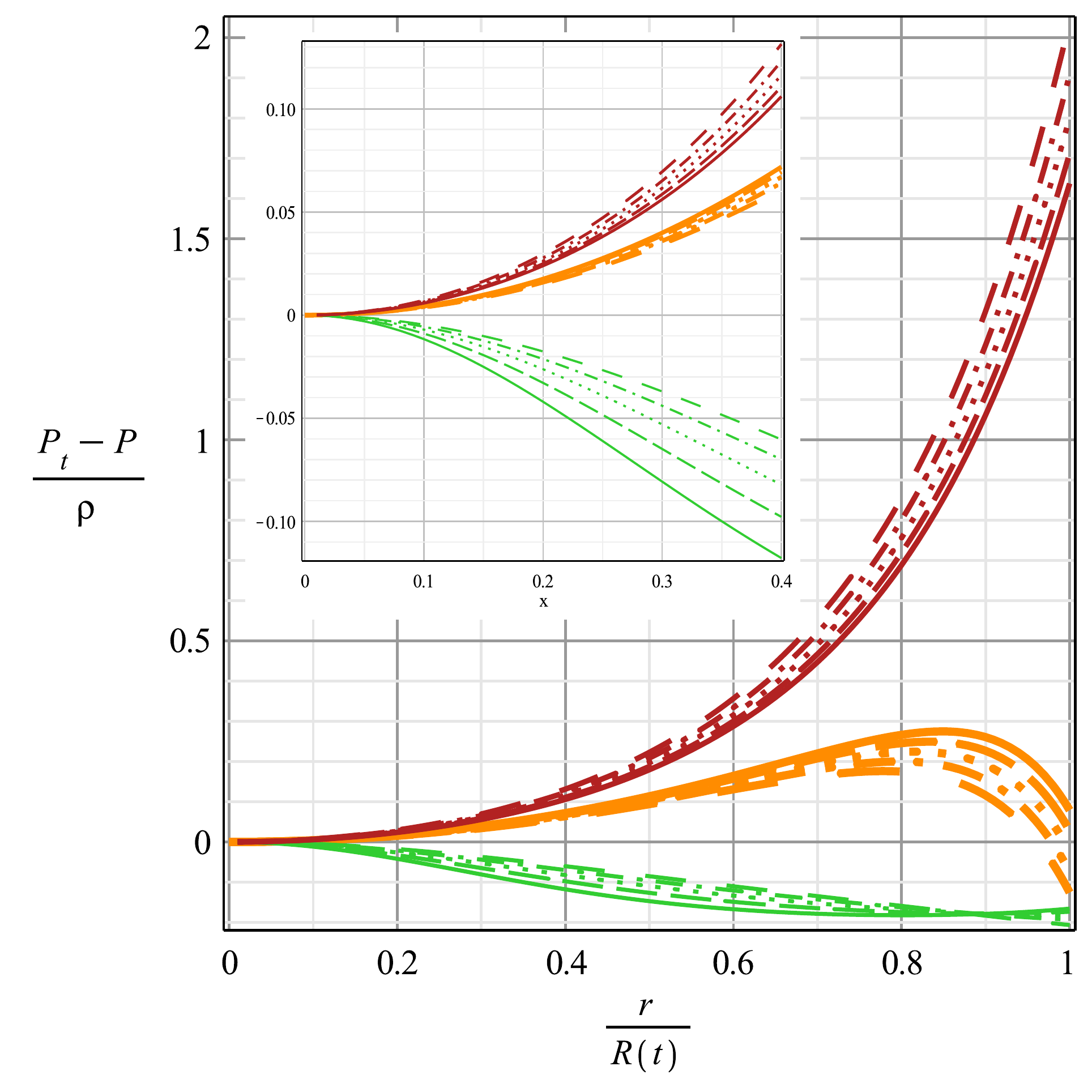}\label{PtFc} }
 
\subfigure[B$\, L_{p}$]
{\includegraphics[scale=0.22]{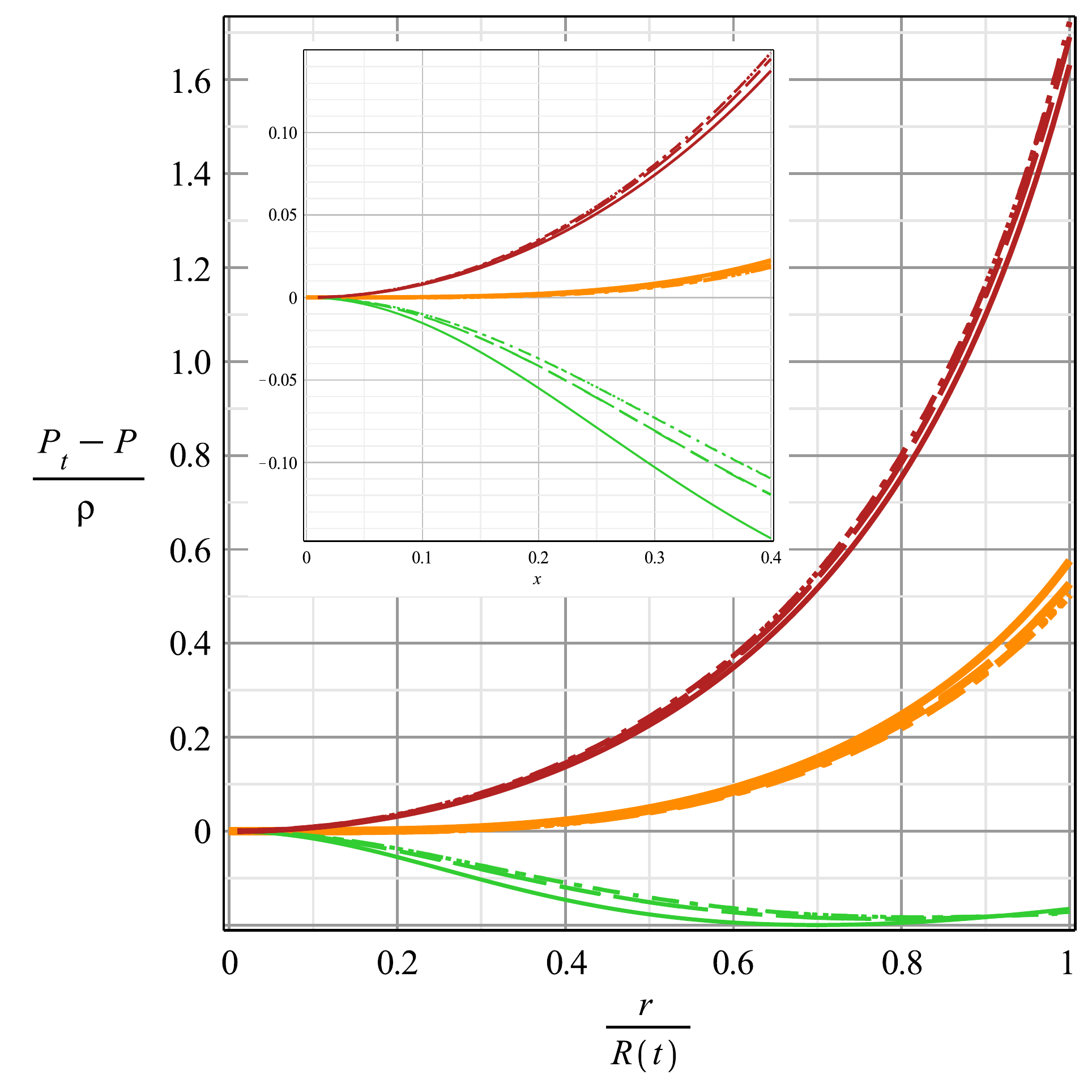}\label{PtBp} }
\subfigure[F$\, L_{p}$]
{\includegraphics[scale=0.22]{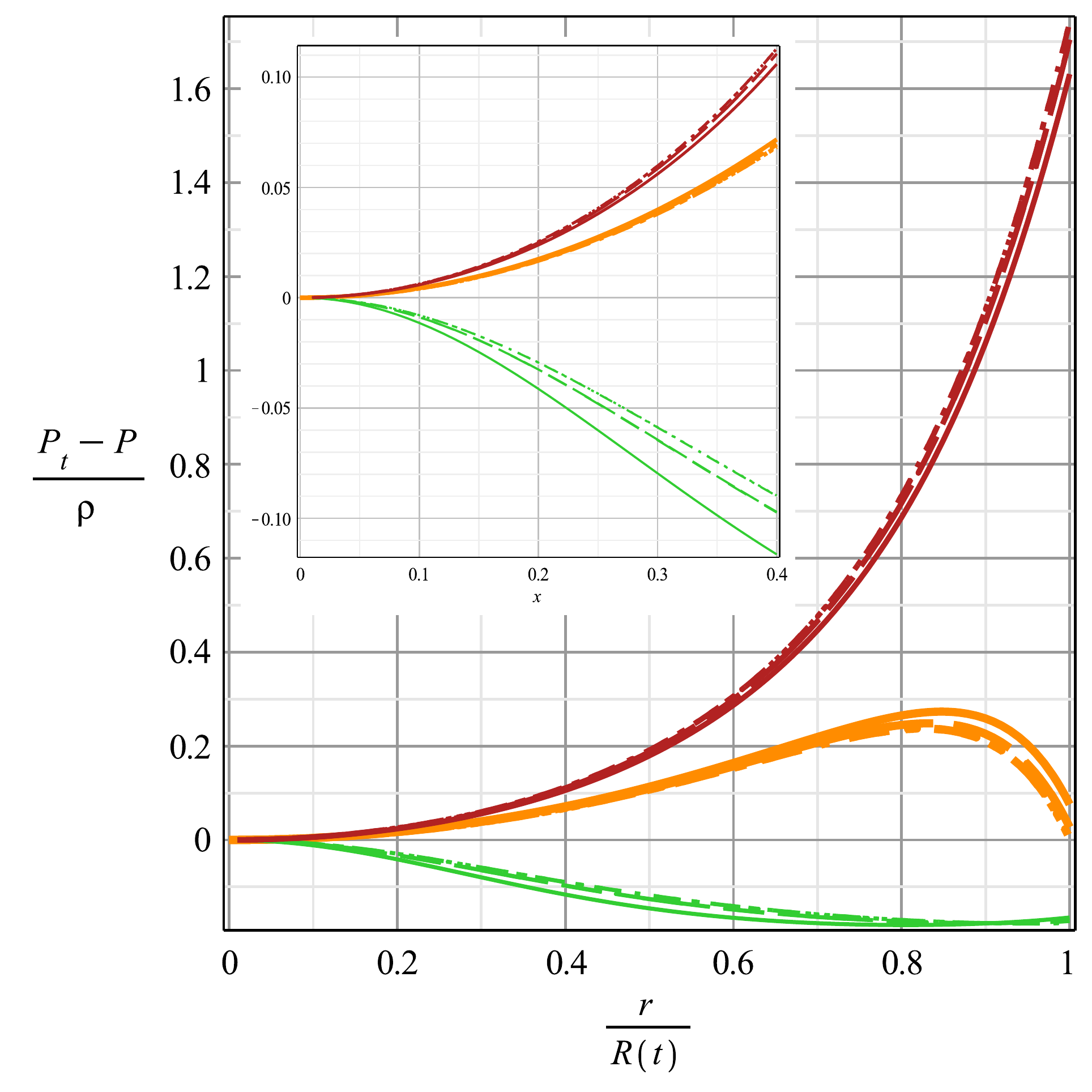}\label{PtFp}} 
\caption{Profiles for $\frac{P_{\perp}-P}{\rho}$ for models having $M_{0}=2M_{\odot}$. Plates \ref{PtBc} and \ref{PtFc} stand for constant luminosity, while \ref{PtBp} and \ref{PtFp} represent those with a pulse-like one. Both \textit{harder} anisotropic models present an increase of the anisotropy for the outer layers and it also increases with time. Concerning the QLEoS models it strongly depend of the selected density profile and have a qualitatively different evolution with time. In the case of \textit{softer} anisotropic EoS it shows a maximum (biggest pressure difference) at the inner layers and then decreases as it approaches to the surface of the distribution.}
\end{figure}

Evolving profiles of dimensionless physical variables, $\rho(r,t)/\rho_{c}(0)$, $P(r,t)/P_{c}(0)$ and the proportionality index $\mathcal{I}_{A}=\frac{P_{\perp}(r,t)-P(r,t)}{\rho(r,t)}$, are shown in Figures \ref{rhoBc}-\ref{PtFp}. 

It is clear from figures \ref{rhoBc}-\ref{rhoFp} that we are using (almost) the same density profile for our LEoS (\textit{hard} \& \textit{soft}) and QLEoS modelling. From Figures \ref{PBc}-\ref{PFp} the distinction between a \textit{hard} ($h=11/10$) and \textit{soft} ($h=1/80$) LEoS can be observed. Pressure gradient distribution, $\mathrm{d} P(r,t)/\mathrm{d}r$ for QLEoS models are  between the corresponding gradients for \textit{hard} \& \textit{soft} anisotropic LEoS.  

The evolution of proportionality indexes for each model are displayed in Figures \ref{PtBc}-\ref{PtFp}. This particular index, in the isotropic case, $\mathcal{I}_{I}=P/\rho$, could represent a family of simple equation of state, i.e. $P~\propto~\rho$. For quasi-local anisotropic fluids, $\mathcal{I}_{A}$ points out where the the fluid could be associated with a \textit{hard} or \textit{soft} EoS. Additionally, the proportionality index could also provide an idea of the compressibility of the dissipative fluid at each point. 
Notice that for QLEoS models it strongly depends on the selected density profile. Both \textit{harder} anisotropic models present an increment (increasing with time) of the anisotropy at outer layers. Concerning the QLEoS models the $\mathcal{I}_{A}$ profiles are heavily conditioned by the selected density distribution and have qualitatively different evolutions in time. For these quasi-local configurations, $\mathcal{I}_{A}$ diminishes with time. In the case of \textit{softer} anisotropic EoS models, $\mathcal{I}_{A}$ has a maximum (biggest pressure difference) at the inner layers and then decreases as it approaches to the surface of the distribution. 

\begin{figure}[h!]
\centering
\includegraphics[scale=0.31]{V60CaptionAll.pdf} \\
\subfigure[2MB$\, L_{c}$]
  {\includegraphics[scale=0.18]{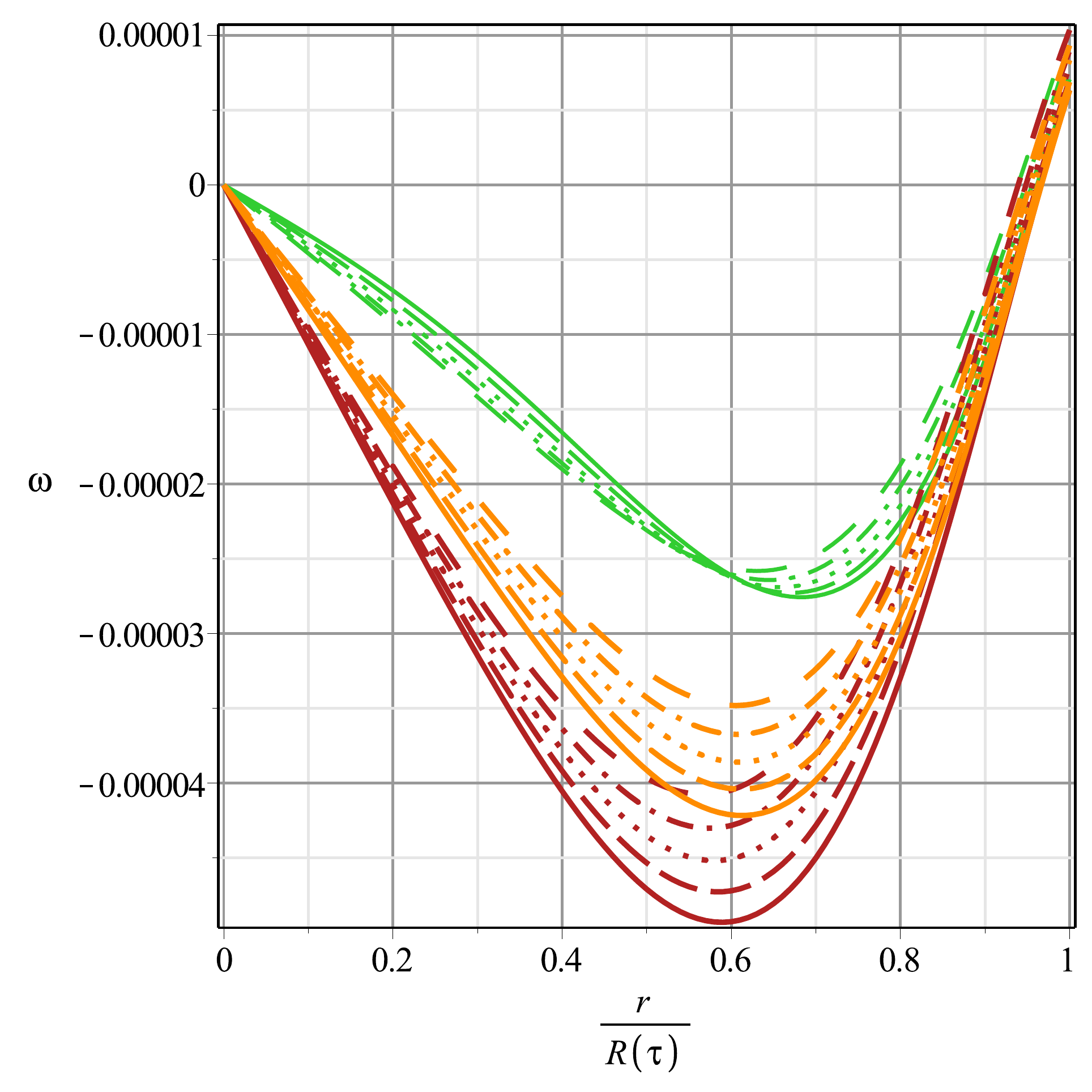} \label{omegaBc}}
\subfigure[20MB$\, L_{c}$]  
 {\includegraphics[scale=0.18]{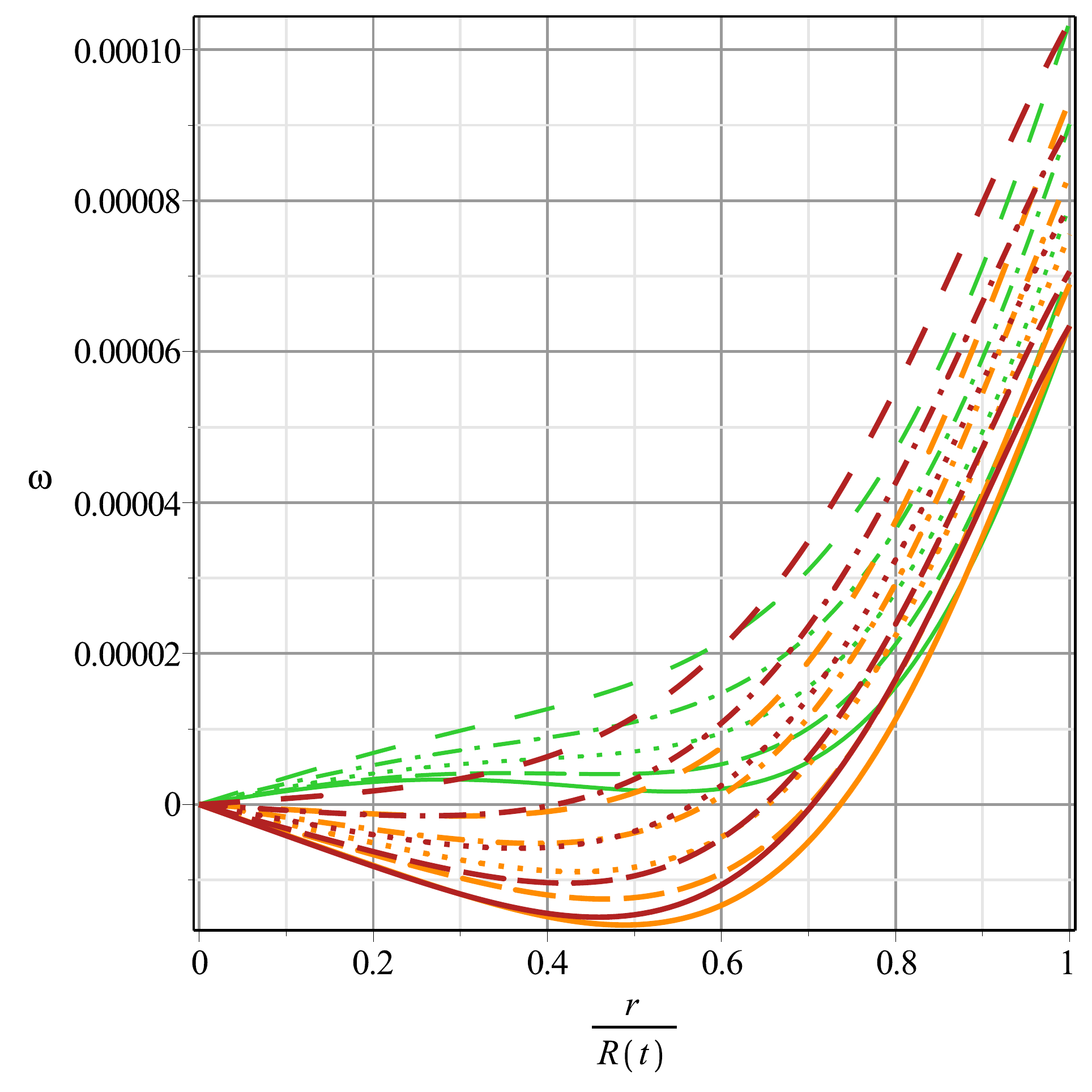} \label{omegaMc20M}}
\subfigure[2MF$\, L_{c}$]
{\includegraphics[scale=0.18]{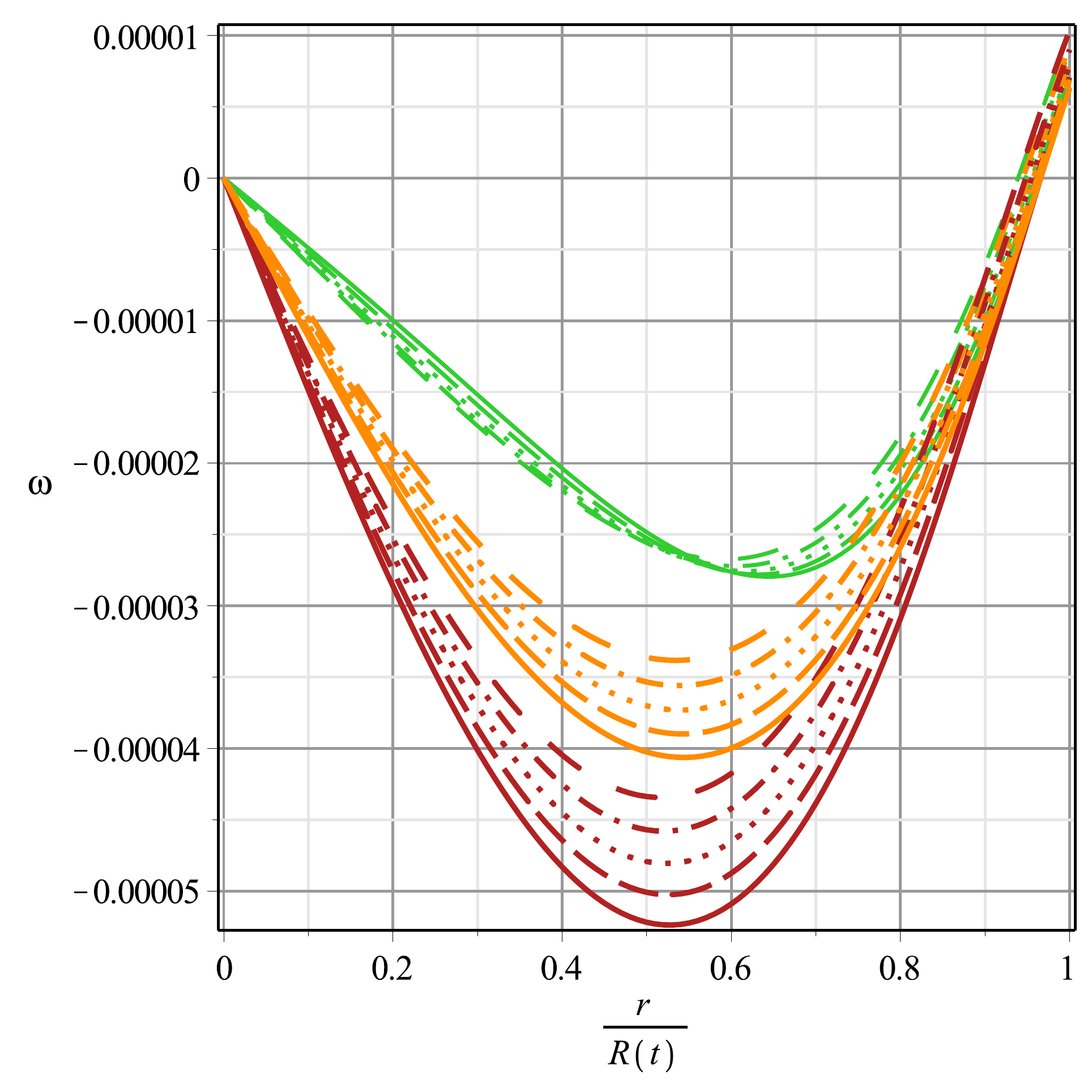} \label{omegaFc}}
\subfigure[20MF$\, L_{c}$]
{\includegraphics[scale=0.18]{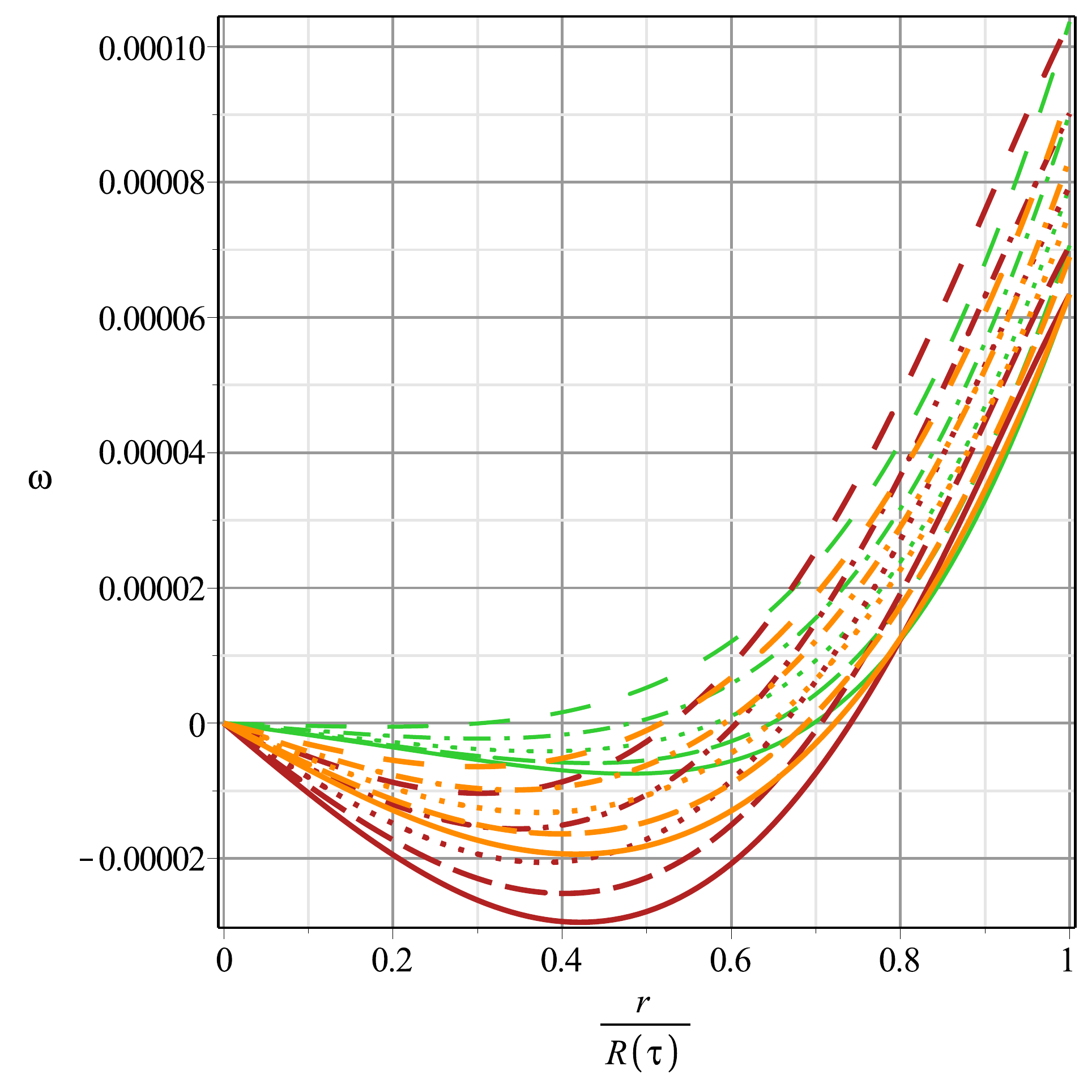} \label{omegaFc20M}}

\subfigure[2MB$\, L_{p}$]
{\includegraphics[scale=0.18]{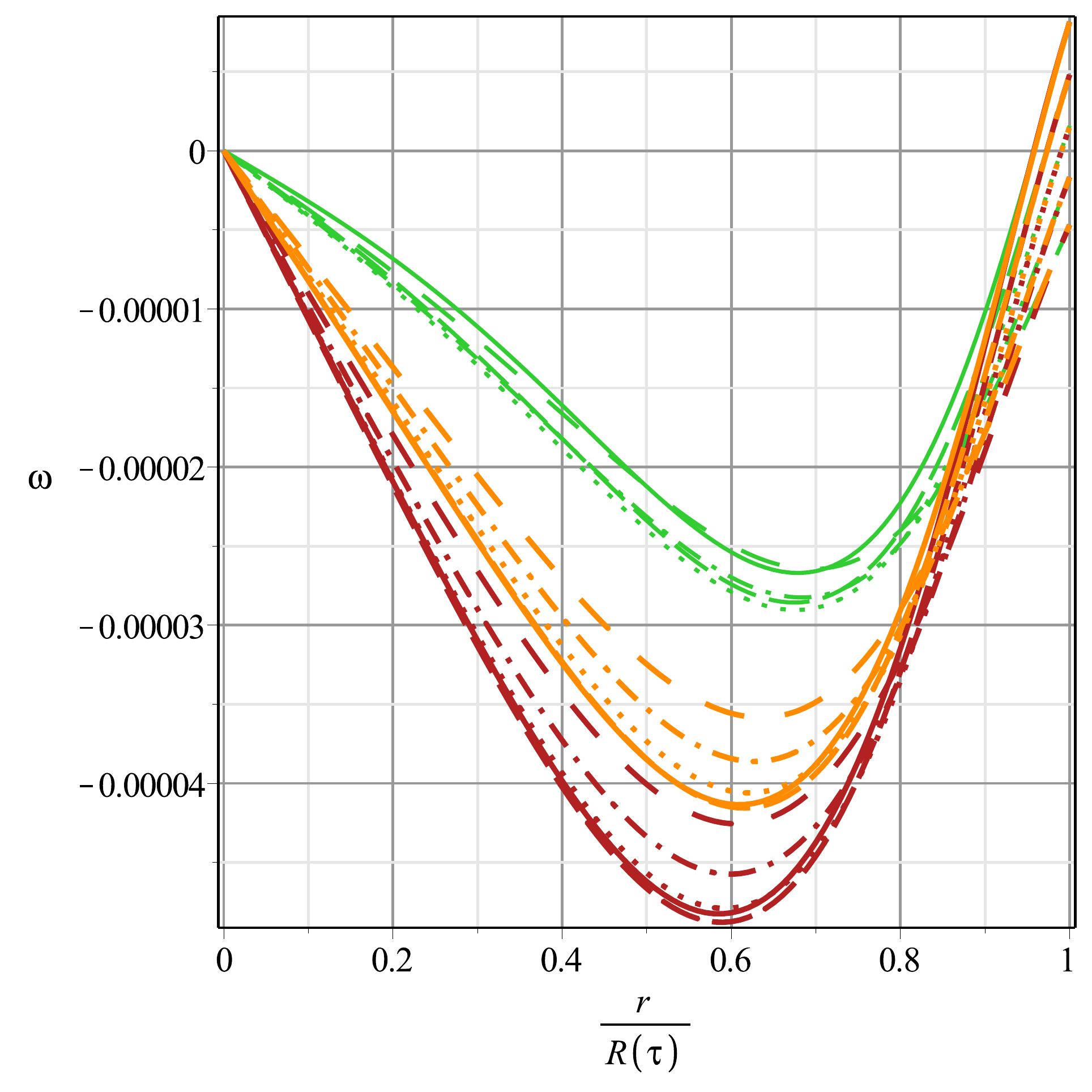} \label{omegaBp}}
\subfigure[20MB$\, L_{p}$]
{\includegraphics[scale=0.18]{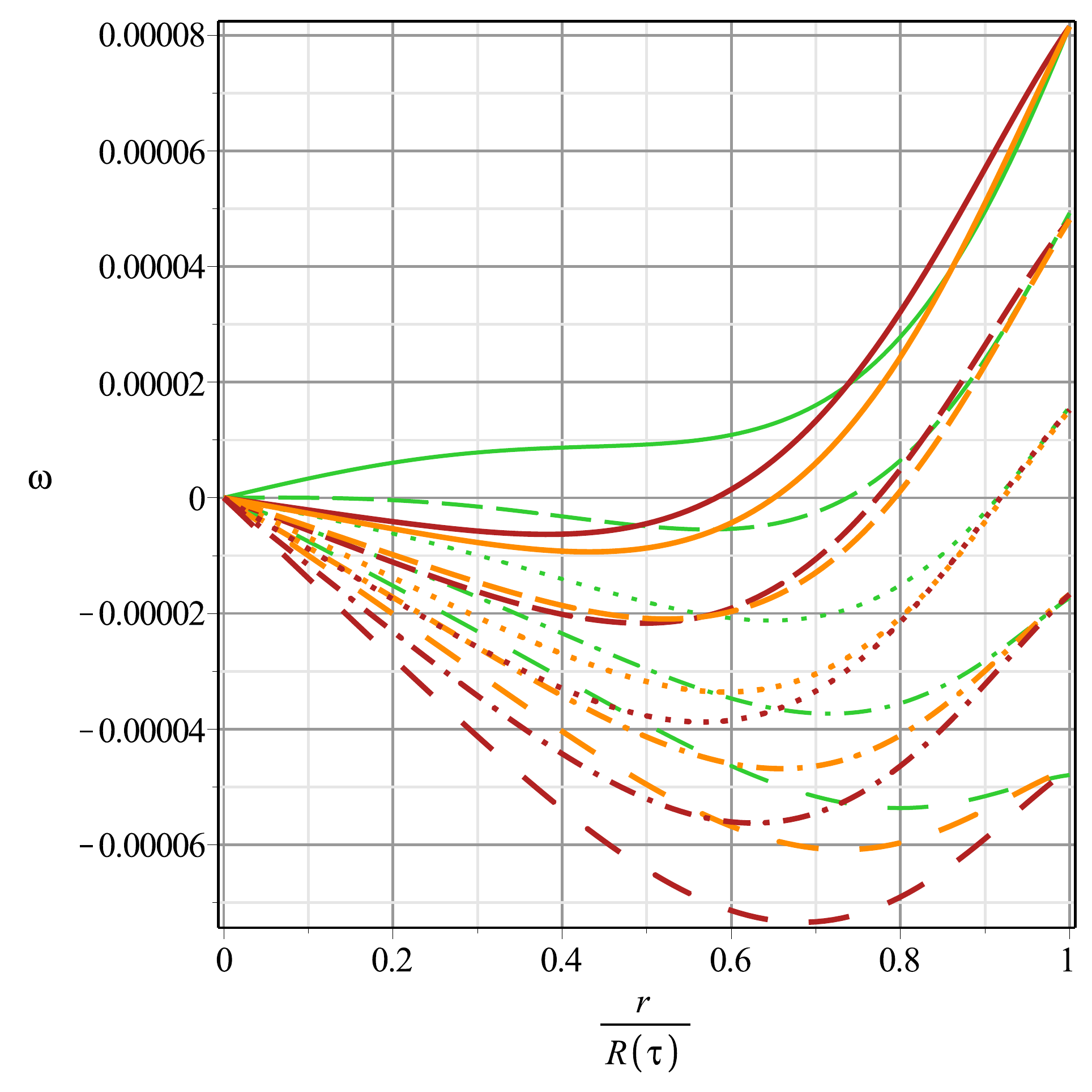} \label{omegaBp20M}}
\subfigure[2MF$\, L_{p}$]
{\includegraphics[scale=0.18]{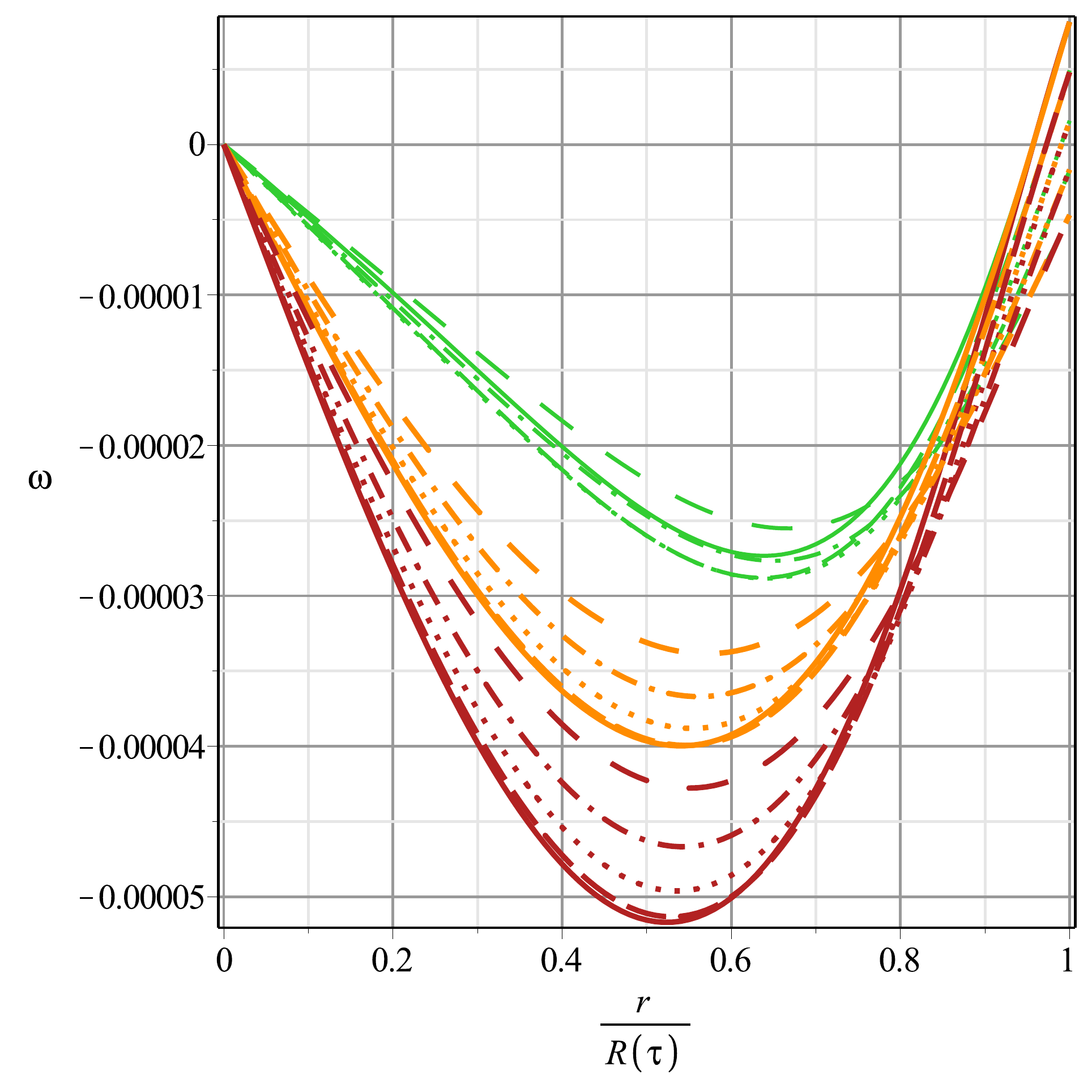} \label{omegaFp}}
\subfigure[20MF$\, L_{p}$]
{\includegraphics[scale=0.18]{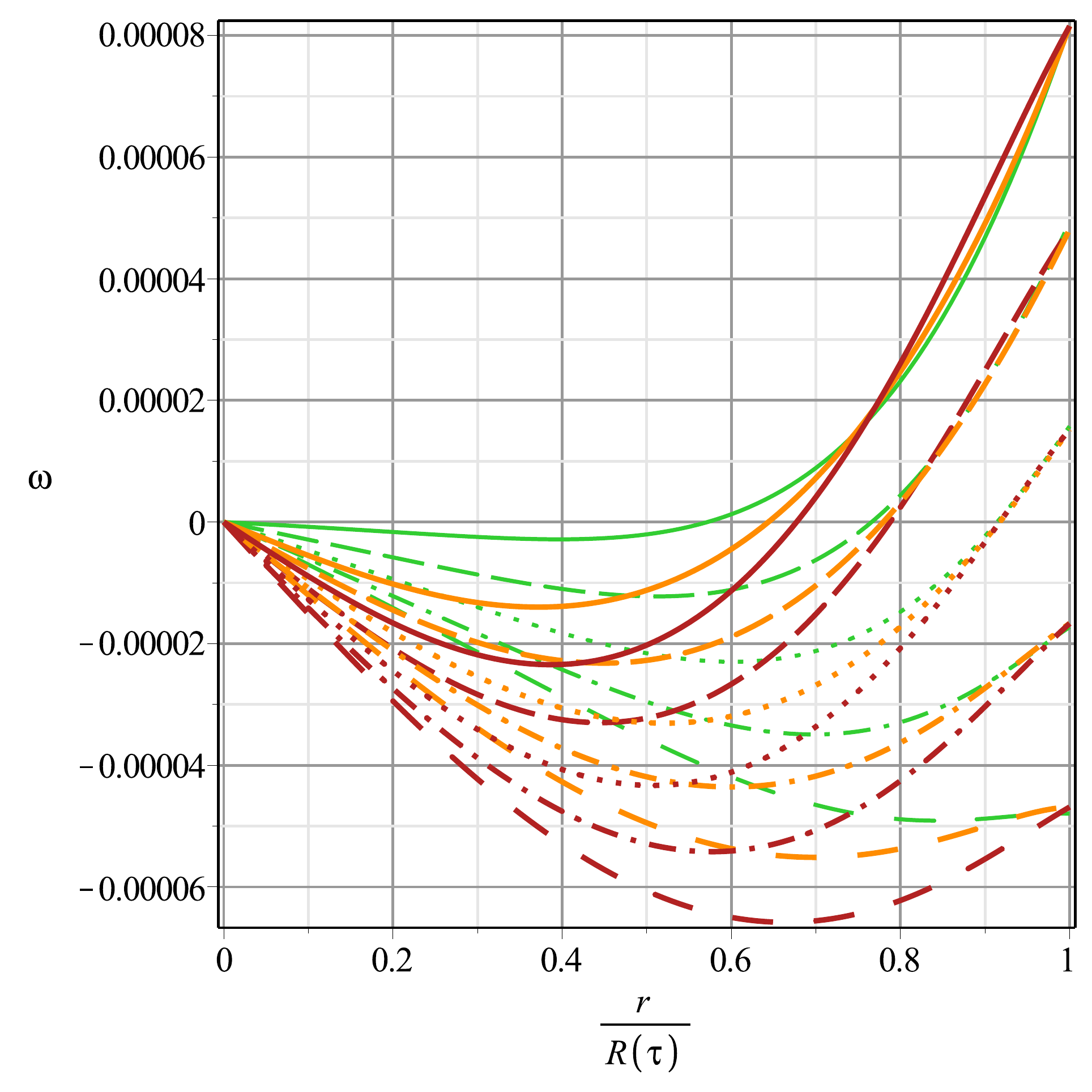} \label{omegaFp20M}}

\caption{Profiles for matter radial velocity, $\omega(r,t)$ for models having $M_{0}=2M_{\odot}$ and $M_{0}=20M_{\odot}$ . Plates \ref{omegaBc}-\ref{omegaFc20M} stand for constant luminosity, while \ref{omegaBp}-\ref{omegaFp20M} represent those with a pulse-like one. The thermal peeling is displayed in all models. It increases for constant luminosity models sketched in plates \ref{omegaBc} and \ref{omegaFc}, but for those with a pulse-like decreasing luminosity represented in plates \ref{omegaBp} and  \ref{omegaFp}, the peeling effect is inverted. With the time evolution all outer mass shells collapse.}
\end{figure}

The thermal peeling can be appreciated in figures \ref{omegaBc} and \ref{omegaFp20M}. It is definitely  present in local and quasi-local configurations for different initial mass configurations ($2M_{\odot}~ \lesssim~M_{0}~\lesssim~20M_{\odot}$).  When a constant luminosity is assumed (figures \ref{omegaBc} through \ref{omegaFc20M}), the peeling increases with time. On the other hand, as it can be appreciated from figures \ref{omegaBp} through \ref{omegaFp20M}, it vanishes for models having pulse-like decreasing luminosity where all the outer layers begin to collapse down. The initial mass of the configuration $M_{0}$ emerges a key parameter to cause the thermal peeling because It increases with increment of the initial mass.

\begin{figure}[h!]
\centering
\includegraphics[scale=0.31]{V60CaptionAll.pdf}  \\
\subfigure[B$\, L_{c}$]
{\includegraphics[scale=0.22]{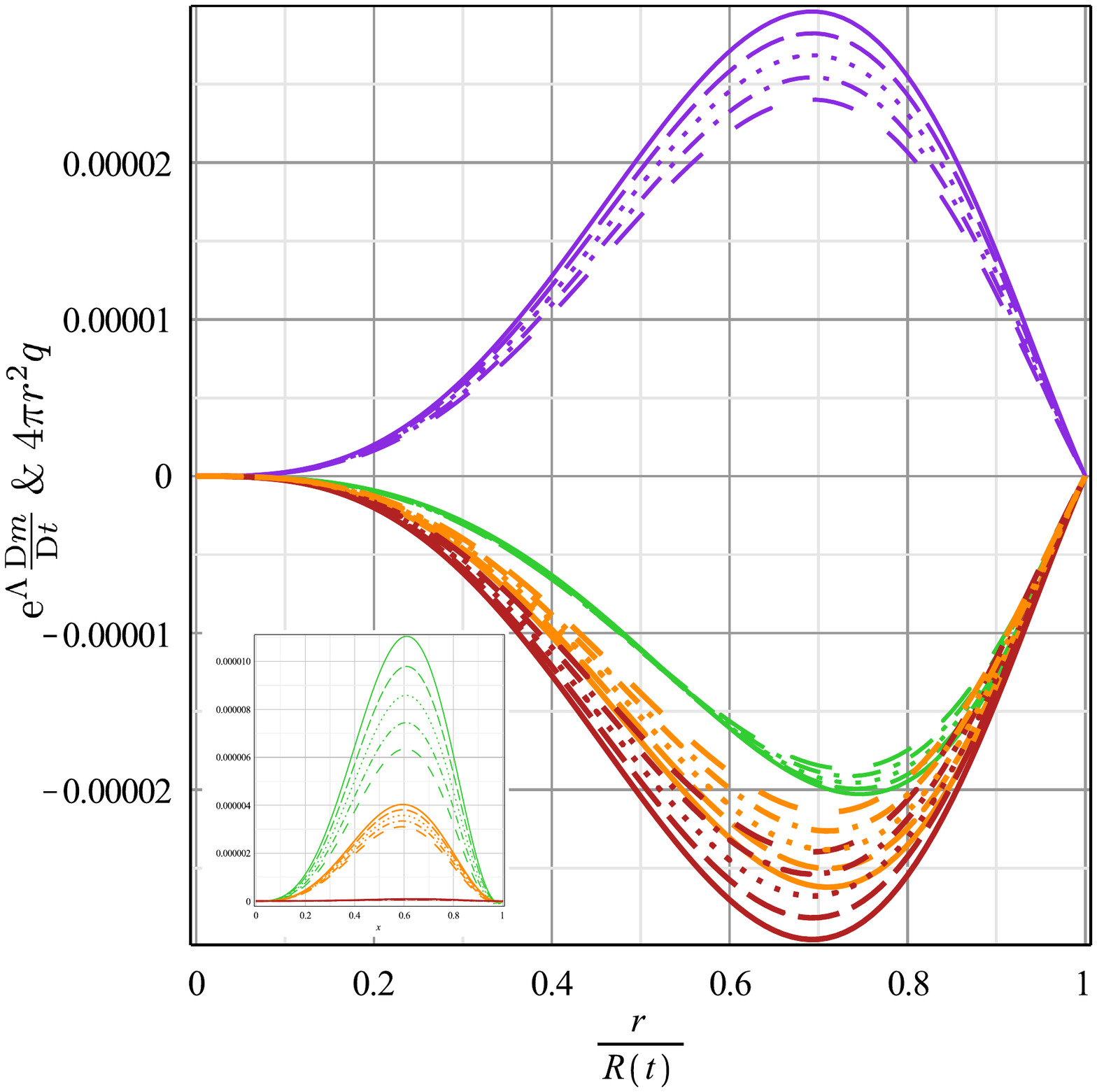} \label{DmMcB}}
\subfigure[F$\, L_{c}$]
{\includegraphics[scale=0.215]{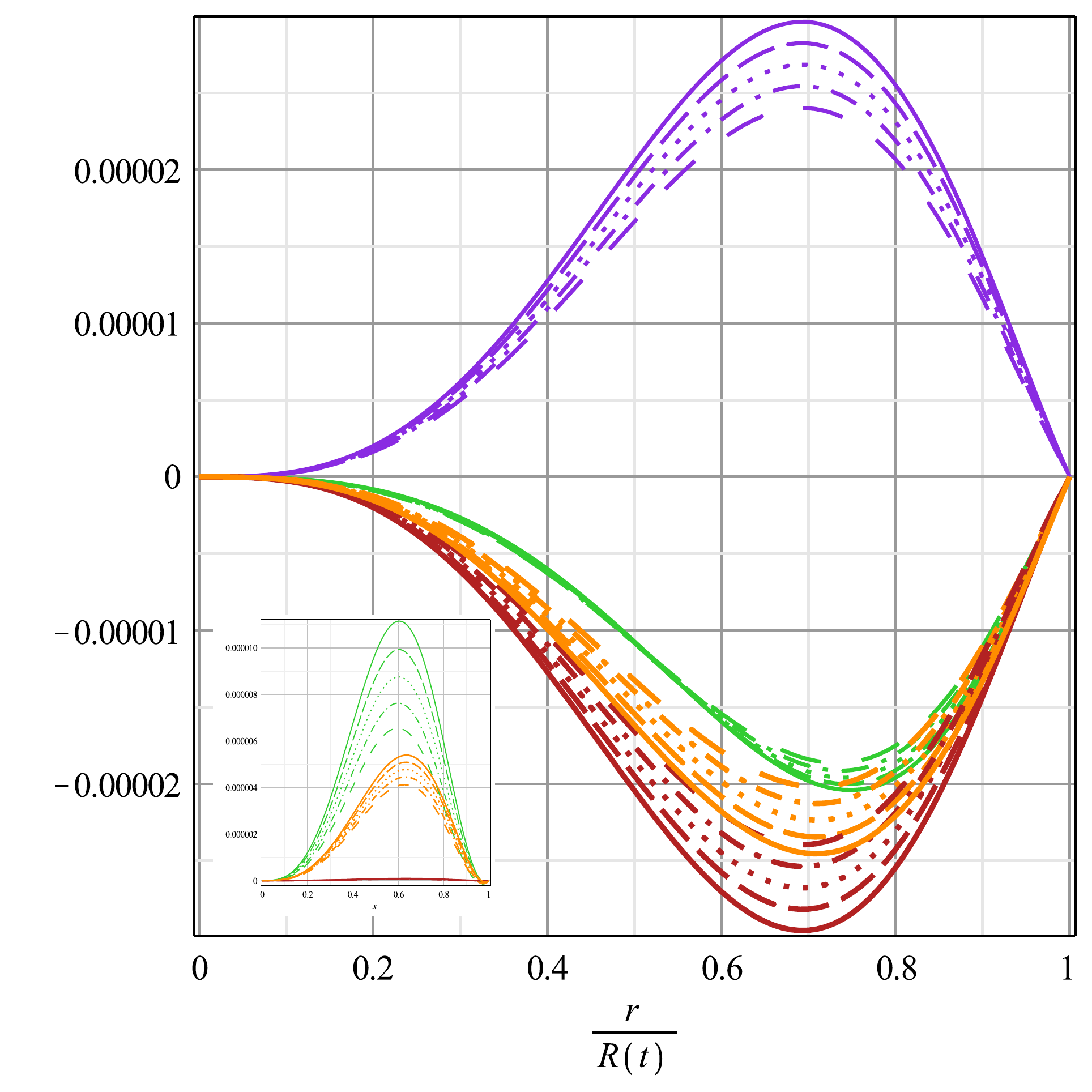} \label{DmMcF}}

\subfigure[B$\, L_{p}$]
{\includegraphics[scale=0.22]{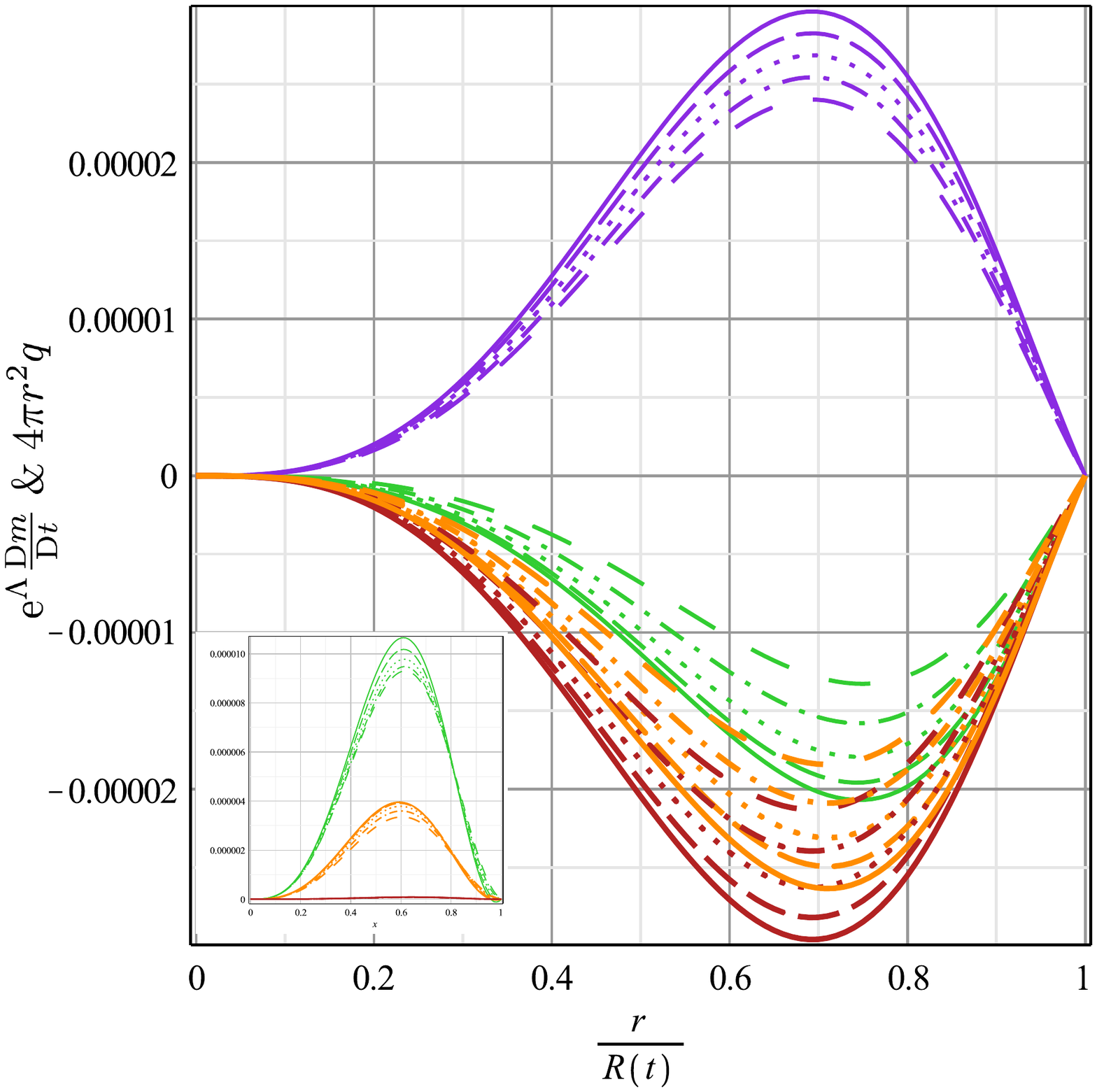} \label{DmMpB}}
\subfigure[F$\, L_{p}$]
{\includegraphics[scale=0.215]{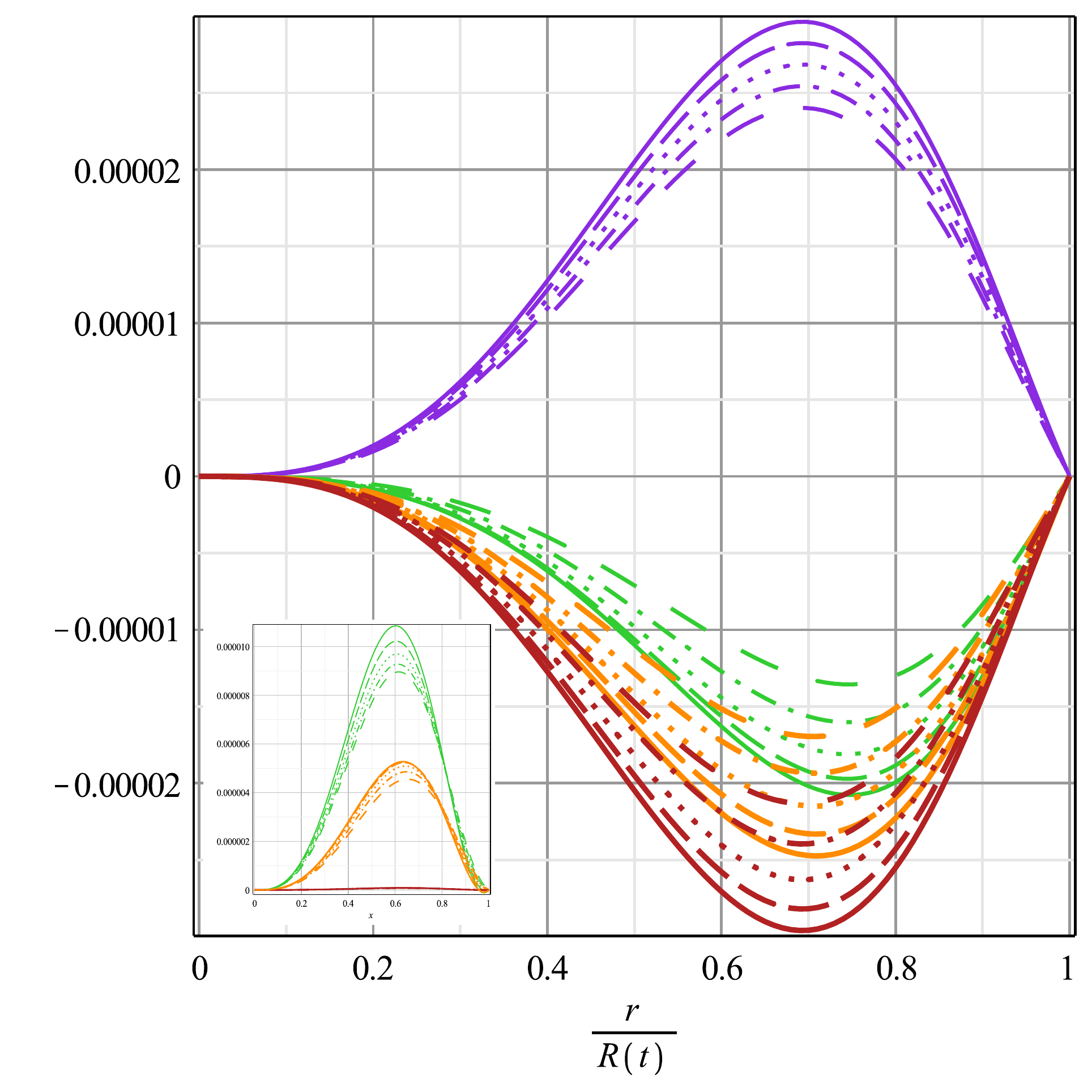} \label{DmMpF}}
 
\caption{Contribution to the matter velocity, $\omega$, for models having $M_{0}=2M_{\odot}$.  Radiation field, $4 \pi r^2 q$, is represented by purple lines.  Green, yellow and red lines are for convective mass $\mathrm{e}^{\Lambda}\frac{\mathrm{D}m}{\mathrm{D}t}$, with $\Lambda =\lambda -\nu$ for LEoS and $\Lambda = -K(t)$ QLEoS. Plates \ref{DmMcB} and \ref{DmMcF} stand for constant luminosity, while \ref{DmMpB} and \ref{DmMpF} represent pulse-like ones. These figures emphasise  the peeling as a purely thermal effect and the need to have an intense radiation filed to be present. The inner plots representing the sum of both contributions: $\mathrm{e}^{\Lambda}\frac{\mathrm{D}m}{\mathrm{D}t} + 4 \pi r^2 q$, illustrating how important is the radiation field to generate peeling.}
\end{figure}

Figures  \ref{DmMcB} through \ref{DmMpF}  display the evolution of the two terms that contribute to the matter velocity for local and quasi-local models (equations (\ref{ConvectiveDerivmlocal}) and (\ref{SignOmega})). As it is clear from these figures peeling is a purely thermal effect due to the intense radiation filed. At each shell $r=constant$, the mass is diminishing while the radiation flux is increasing. The inner plots, representing the addition of both contributions, i.e. $\frac{\mathrm{e}^{\Lambda}}{4\pi r^{2}}\frac{\mathrm{D}m}{\mathrm{D}t} + 4 \pi r^2 q$, illustrate the phenomena of expelling the outer mass shells. Observe that we have represented  $\Lambda =\lambda -\nu$ for local anisotropic models and $\Lambda = -K(t)$ for quasi-local matter distributions.

\begin{figure}[h!]
\centering
\includegraphics[scale=0.31]{V60CaptionAll.pdf}  \\
\subfigure[2MB$\, L_{c}$] 
{\includegraphics[scale=0.22]{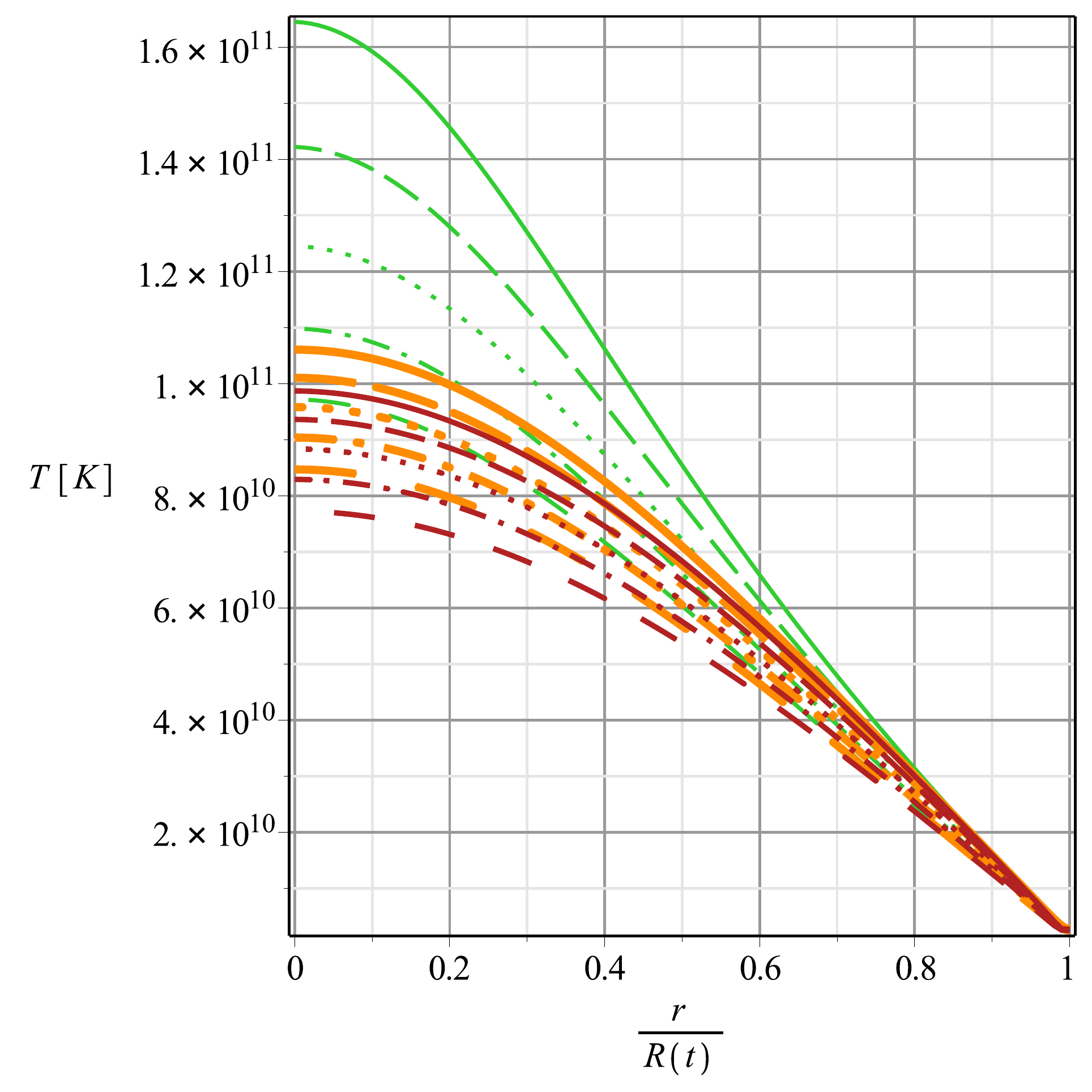} \label{thermalprofileBc}}
\subfigure[2MF$\, L_{c}$]
{\includegraphics[scale=0.22]{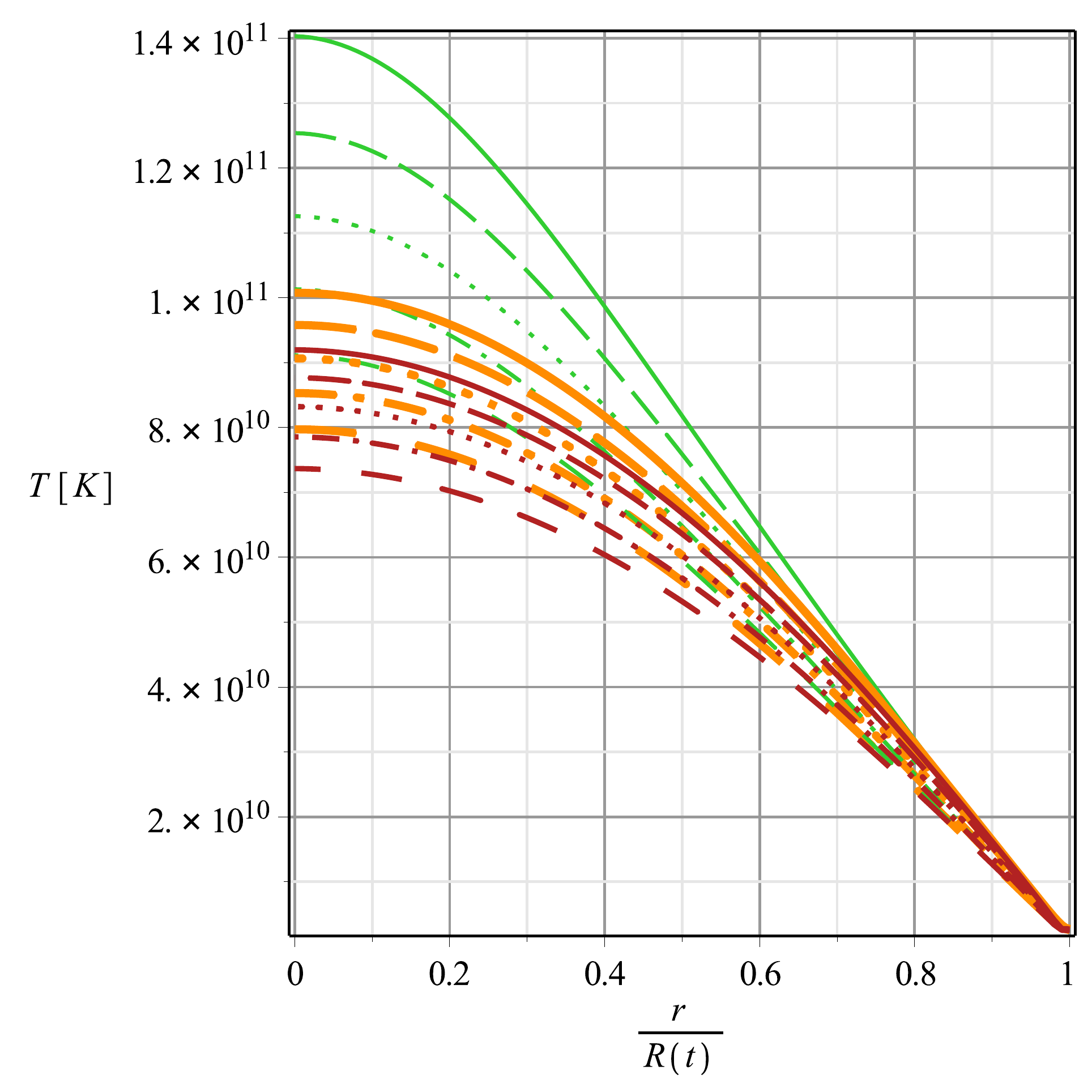} \label{thermalprofileFc}}

\subfigure[2MB$\, L_{p}$]
{\includegraphics[scale=0.22]{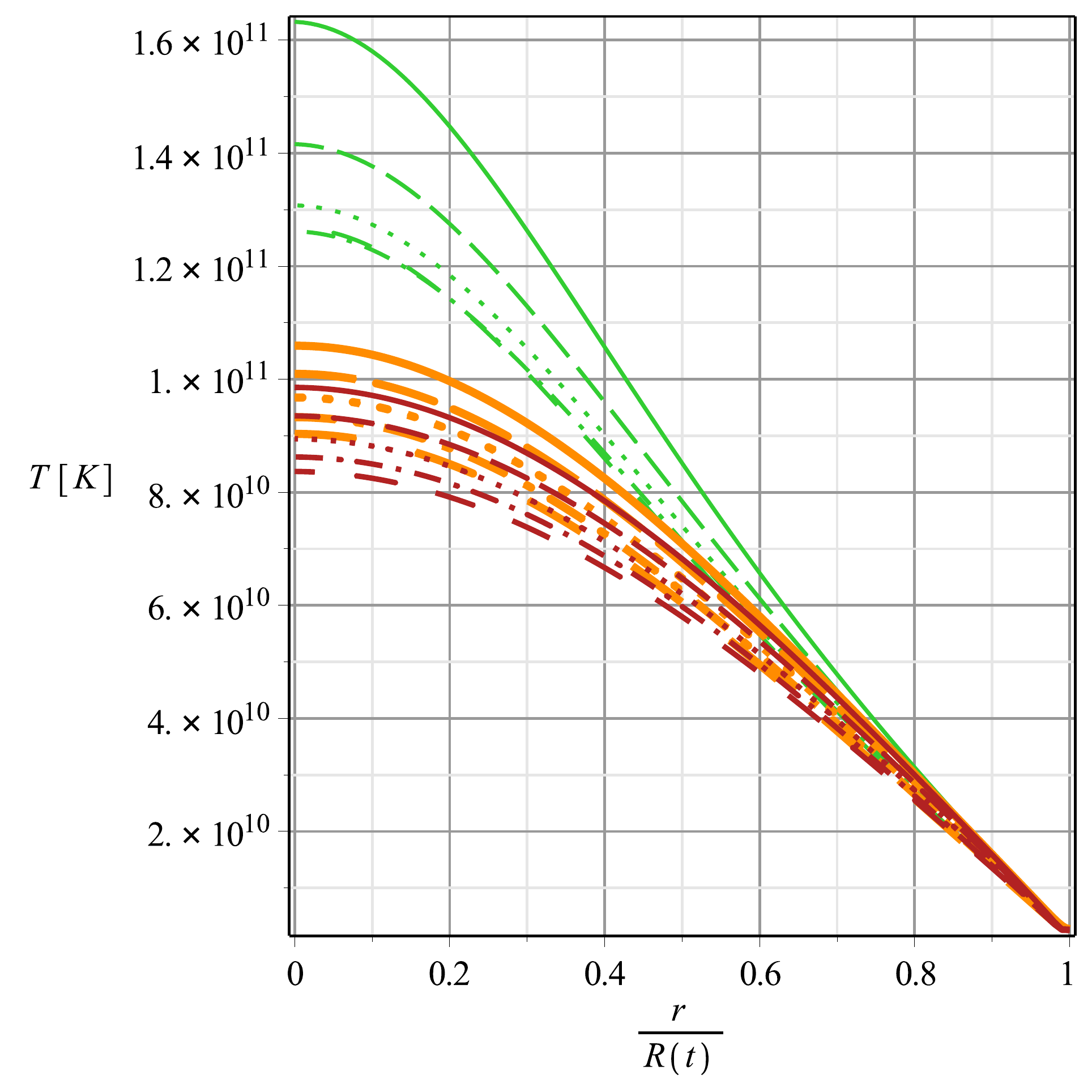} \label{thermalprofileBp}}
\subfigure[2MF$\, L_{p}$]
{\includegraphics[scale=0.22]{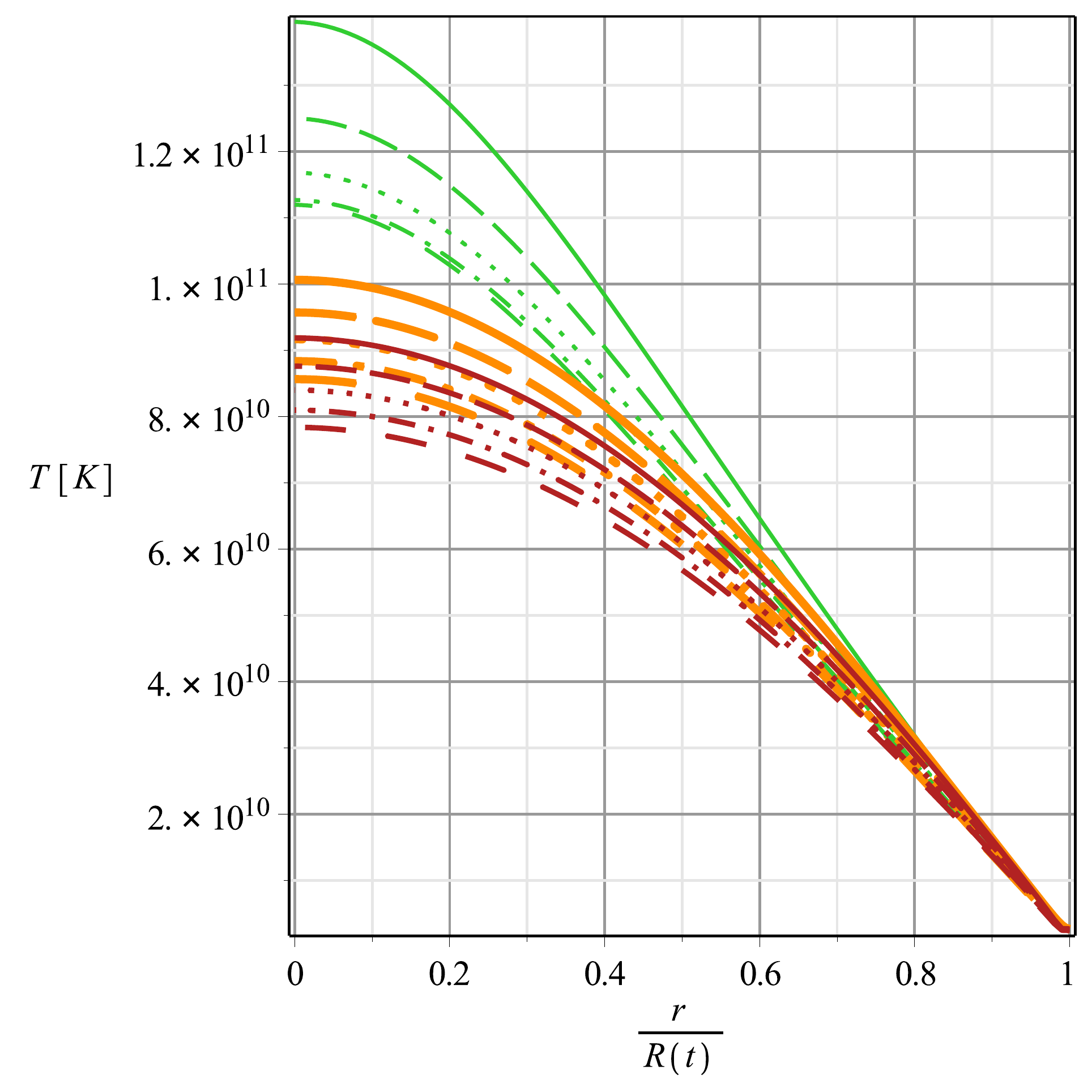} \label{thermalprofileFp}}
\caption{Profiles for Temperature distribution, $T(r,t)$,  emerging from a causal thermodynamics description within the matter distribution.  Plates \ref{thermalprofileBc} and \ref{thermalprofileFc} stand for constant luminosity, while \ref{thermalprofileBp} and \ref{thermalprofileFp} represent those with a pulse-like one. Quasi local Bushdalh configurations present lower temperatures distributions than the corresponding local ones. But when FGM-models are consiered the temperature depends on the shape of the luminosity.}
 \end{figure}

In figures \ref{thermalprofileBc}-\ref{thermalprofileFp}  temperature distributions are displayed. These distributions emerge from the integration of Cattaneo equation (\ref{SlowCatt2}), for models having $M_{0}~\approx~2M_{\odot}$ and are in the range of $10^{10} K \lesssim T \lesssim 10^{11} K$. In the case of models with $M_{0}~\approx~20M_{\odot}$ we have obtained $6.5 \times 10^{9} K \lesssim T \lesssim 0.9 \times 10^{9} K$. Notice that we have obtained that, greater initial mass could generated cooler models.  Also it is apparent that the degree of anisotropy could have some influence in the thermal behavior of the configurations i.e. \textit{harder} anisotropic LEoS models display higher inner temperatures than \textit{softer} models. Again, QLEoS are in middle of these thermal behaviors. 

\begin{figure}[h!]
\centering
\includegraphics[scale=0.31]{V60CaptionAll.pdf}  \\
\subfigure[2MB$\, L_{c}$]
{\includegraphics[scale=0.22]{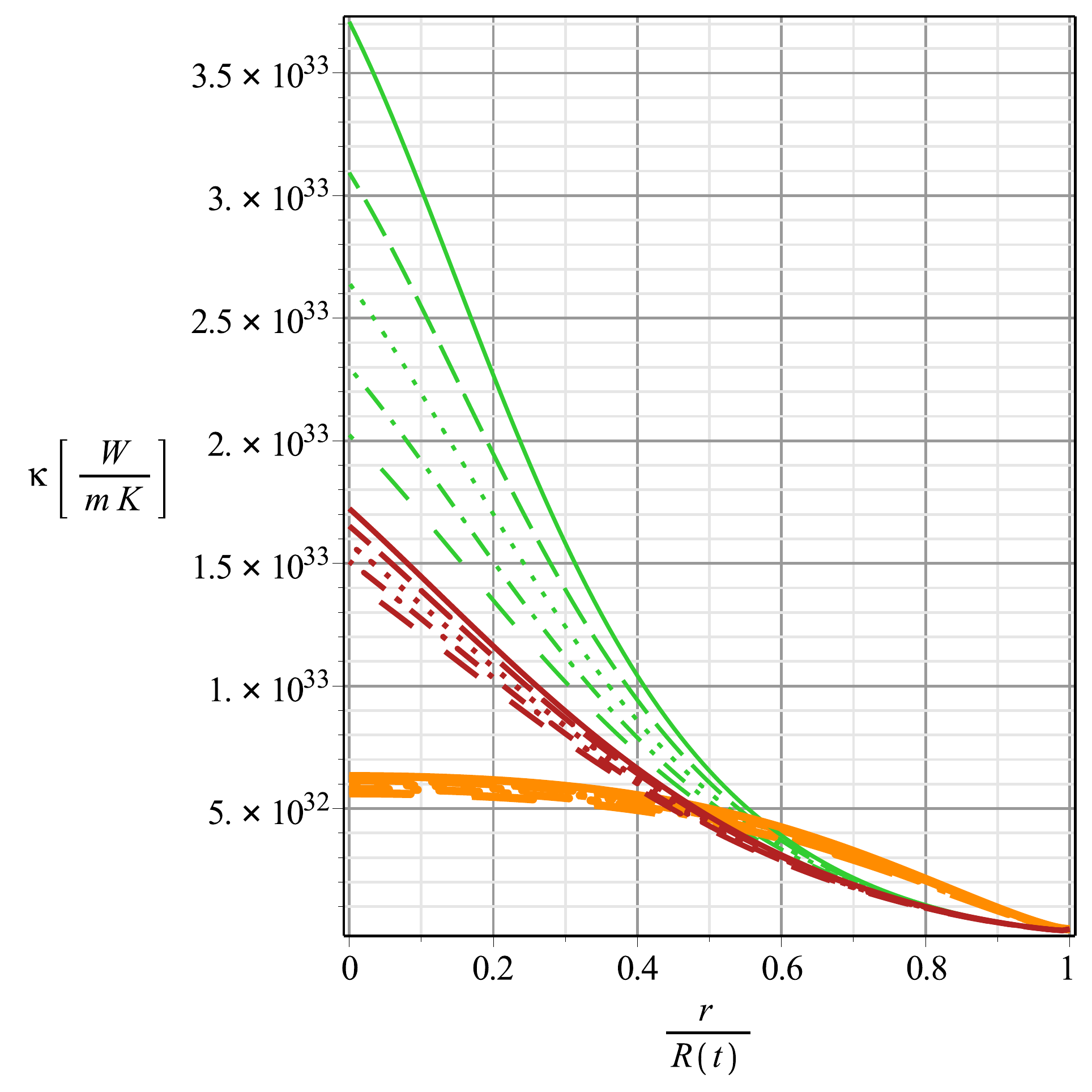} \label{kappaBc}}
\subfigure[2MF$\, L_{c}$]
{\includegraphics[scale=0.22]{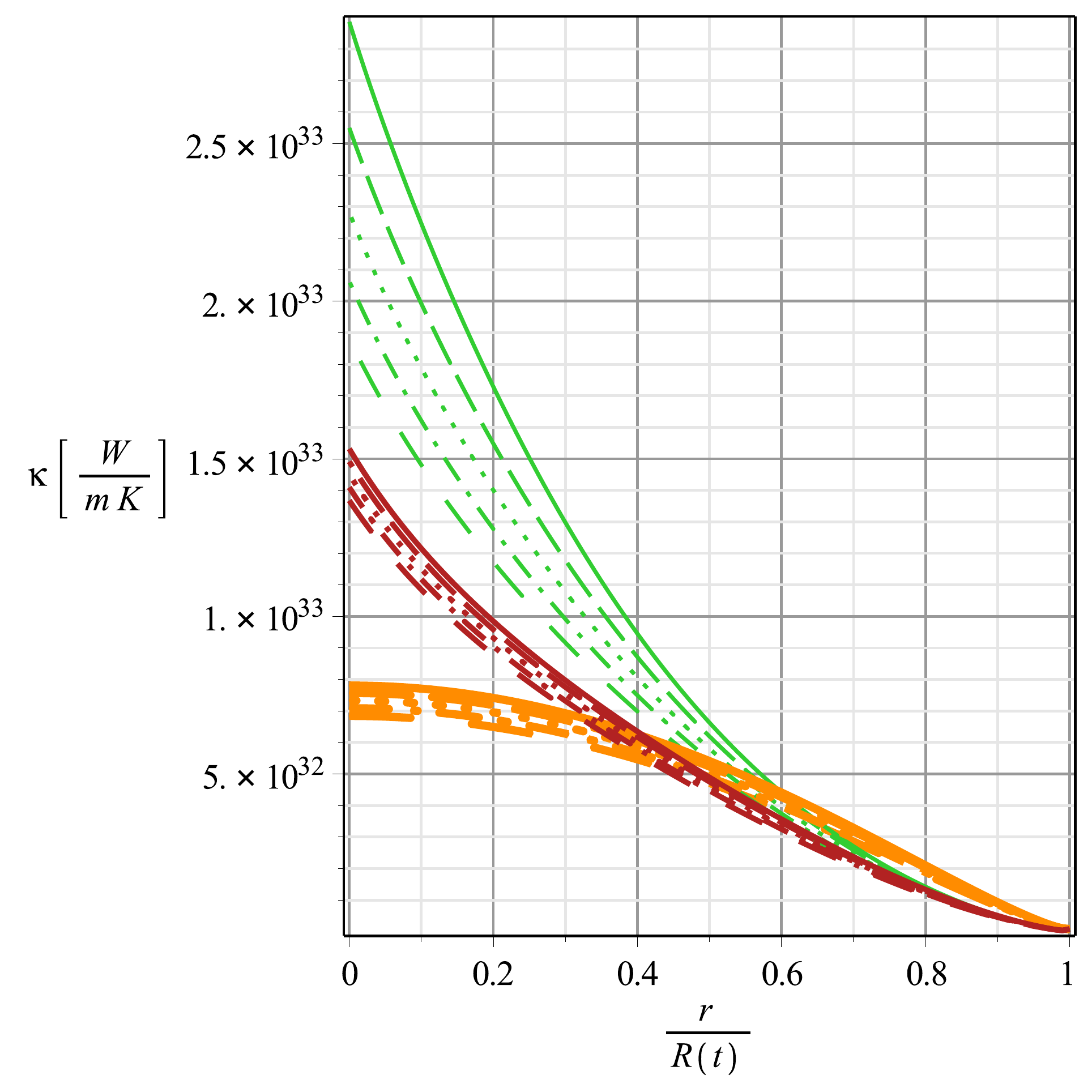} \label{kappaFc}}

\subfigure[2MB$\, L_{p}$]
{\includegraphics[scale=0.22]{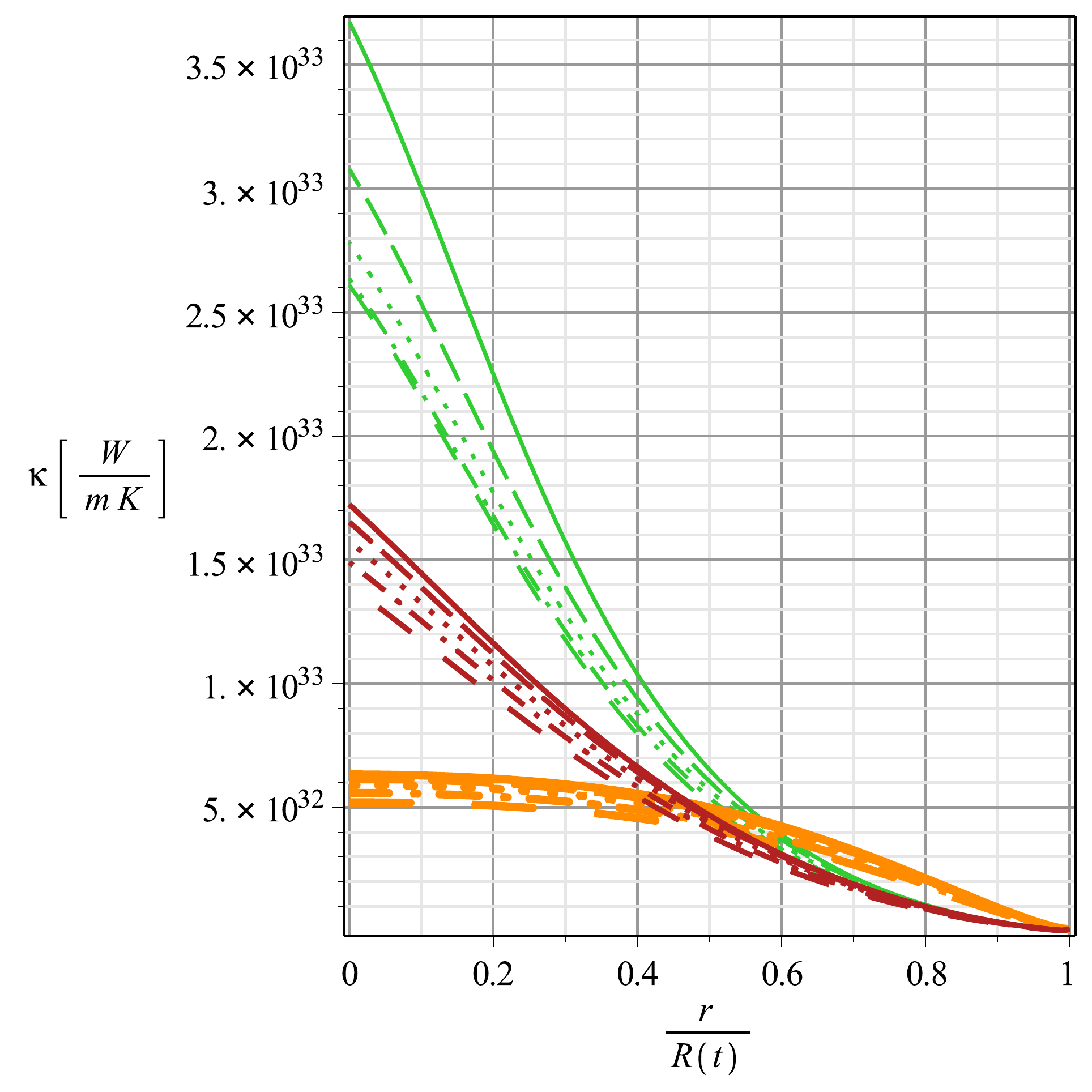} \label{kappaBp}}
\subfigure[2MF$\, L_{p}$]
{\includegraphics[scale=0.22]{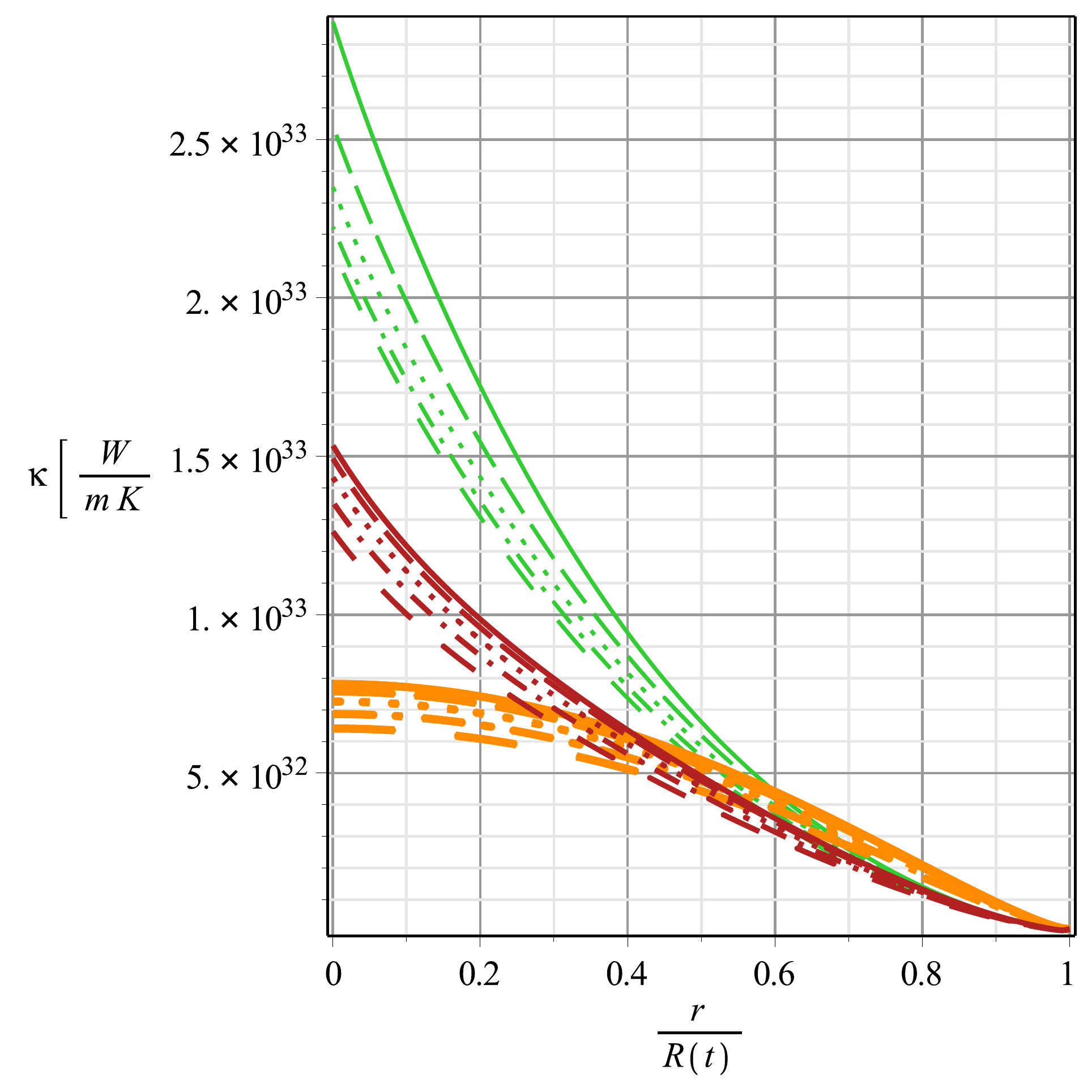} \label{kappaFp}}

 \caption{Profiles for the calculated thermal conductivity.  Plates \ref{kappaBc} and \ref{kappaFc} stand for constant luminosity, while \ref{kappaBp} and \ref{kappaFp} represent those with a pulse-like one.  It seems to be independent of the value of the initial mass of the configuration and QLEoS models display a more homogeneous thermal conductivity than the anisotropic LEoS models.}
 \end{figure}

Finally, figures \ref{kappaBc}-\ref{kappaFp} illustrate the calculated thermal conductivity: $37~\times~ 10^{33}~\lesssim~\kappa~\lesssim~10^{32}$. It seems to be independent of the value of the initial mass of the configuration. QLEoS models display a more homogeneous thermal conductivity than the anisotropic LEoS models. 

The coming section will be devoted to resume the conclusions of this work and we will discuss these similarities.

\section{Some remarks and conclusions}
\label{RemarksConclusions}
We have used a self-consistent frame work to study quasi static gravitational collapse for dissipative anisotropic matter configurations. As we have mentioned before, the quasi-static evolution approximation is not extremely  restrictive because most relevant processes in star interiors take place on time scales that are usually much longer than the hydrostatic time scale. In fact, as we have pointed out, for neutron stars $t_{hyd}~\approx~10^{-4}s$, therefore the thermal peeling we have studied in the present work, occurring for time spam modelling of $t_{mod}~\lesssim~1s$  is fully consistent with this approximation. 

The main objectives of the present work were to study first, under what circumstances it is possible to obtain a thermal peeling in the context of general relativistic anisotropic dissipative matter distributions and secondly, what would be the temperature profiles that could possible  be associated with existence this thermal effect. It was  found that highly relativistic (i.e. $ 0.2 \lesssim~\left(1 -2M_{0}/R_{0}\right)~\lesssim~0.6 $) and highly luminous ($E \approx 10^{45} erg$) local and quasi-local anisotropic matter configurations could exhibit thermal peeling. This effect, appears to be associated with particular extreme astrophysical scenarios, i.e. highly relativistic and very luminous compact object expelling its outer mass shells, evolving quasi-statically.  It was also obtained that the thermal peeling is very sensible to shape of the energy flux profile and to the luminosity emitted by the compact object. Thus, at least for these two simple density profiles for two different types of equations of state (local and quasi-local) the thermal peeling seams not to be associated with early phases of typical compact object cooling because it requires more massive and luminous escenarios. 

Matter distribution are coupled with the Vaidya exterior metric and consistent with the general relativistic dissipative approach, the internal temperature profile is obtained from the fully relativistic second order thermodynamics where the thermal conductivity has been calculated for neutrino diffusion through the ultra-dense matter. Israel-Stewart description is expressed in the quasi-static approximation and the temperature profile is obtained from a Cattaneo-like equation. 

With the parameters shown in  Table \ref{SimulParam} all the models considered fulfill the acceptability conditions for the physical variables:
\begin{equation}
\label{Aceptability1}
\rho(r,t)>0 , \; \; P_{r}(r,t)  \geq 0, \; \; P_{\perp}(r,t) \geq 0, \; \; \mathrm{and} \; \;  \frac{\partial P_{r}(r,t)}{\partial r} \leq 0 \;  .
\end{equation}
Consistent with the quasi-static framework we also have 
$ | \omega(r,t) | \approx  | q(r,t) |  \approx 10^{-5} \lll 1 \; .$
The anisotropy in our modelling is described by an heuristic equation  of state,  (\ref{anisotropi_equation}), which relates the radial and the tangential pressure within the distribution. Despite its simplicity, this equation provides a qualitative description of the relevance of the anisotropy for the structure of relativistic matter distributions. Using this heuristic equation we have considered two types of anisotropic fluids: \textit{hard} LEoS, \textit{soft} LEoS and, additionally we have also study QLEoS. The different comportment of the anisotropy for the \textit{hard} and \textit{soft} LEoS  can be appreciated from Figures \ref{PtBc}-\ref{PtFp}. Again we would like to stress  that we are not describing the microscopic origin of the local  anisotropy. As it was extensively discussed in reference \cite{HerreraSantos1997}, it could emerge from a variety of physical scenarios and plays important roles at different scales on several astrophysical objects. 

 Again, we have to mention that this scenario resembles a previously reported picture for a conformally flat quasi-local model \cite{MunozNunez2006}.   Models emerging from  Buchdahl and Florides-Gokhroo-Mehra static density profiles, present this thermal effect when most of the radiated energy comes from the mid layers of the distribution.  As it can be appreciated form figures \ref{omegaBc}-\ref{omegaFp20M} and 
 \ref{DmMcB}-\ref{DmMpF} the leading term is the radiation flux which is responsible for the positive sign of the matter velocity when the peeling occur.  The thermal energy generated in the middle of the  mass the distribution is transformed into kinetic energy that generates the ejection of the outer  layers. From our modelling we have obtained this conclusion either, by providing a peeling velocity profile and obtaining a radiation flux distribution with a maximum centered at the middle of the matter distribution or by given a radiation profile where most of the radiated energy comes from the mid layers and obtaining the peeling phenomena for both local and quasi-local models. This particular radiation flux profile was also obtained in the case of a quasi-local gravitational collapse with no approximation\cite{HernandezNunezPercoco1999}. We would like to stress that we are not intending to describe the microphysics responsible of the dissipation within the matter configuration, instead we are pointing out a possible relation between this particular energy flux profile and the thermal peeling effect. Additionally, it is obtained that if the thermal peeling occurs, it will happen under very particular circumstances concerning, no only for an specific energy flux profile, but also it is very sensible to the shape of the luminosity. This is evident when we compare the evolution of the radial velocity in Fig \ref{omegaBc}-\ref{omegaFc20M} with those in \ref{omegaBp}-\ref{omegaFp20M} where the peeling is ``inverted'' and the shells  that were initially expanding collapse down. The above mentioned picture for the thermal peeling has been encountered for an ample range of the initial mass of the matter configuration, i.e. $2M_{\odot} \lesssim M_{0} \lesssim 20M_{\odot}$. 

It seems that the temperature profile is very sensible of the anisotropic equation of state. \textit{Harder} anisotropic equations of state ($P_\perp < P$) lead to hotter matter configurations with higher thermal conductivity. Finally it appears that QLEoS leads to models with more homogeneous thermal conductivity.

At it was stressed before, it seems to be useful to consider relatively simple and consistent relativistic modelling having the key physics constituents, i.e. General Relativity, dissipation and a plausible EoS, then analyze the essential features emerging from this framework which could be an  arena for the evolving numerical codes \& simulation  environments. In this context, despite its attractiveness to describe a compact object expelling its outer mass shells thermal, the peeling appears to be associated with very extreme astrophysical scenarios.

\section*{Acknowledgment}
We are grateful for the financial support of the Vicerrectoría de Investigación y Extensión de la Universidad Industrial de Santander. One of us, LB, gratefully acknowledge the financial support of Colciencias, under the program of Jóvenes Investigadores. LB has also been profited by the hospitality and fruitful discussions and financial support at the International Center for Relativistic Astrophysics, Roma-Italy. LAN has been benefited with the always fresh, occurrent and enlightening morning discussion with L. Herrera.

\newpage
\newpage

\newpage

\newpage

\newpage

\newpage

\newpage
 
\newpage

\end{document}